%% file: substBasedFoundation.tex
\documentclass{llncs}

\usepackage{etex}

\usepackage[all,cmtip,2cell]{xy}

\usepackage{booktabs}   %% For formal tables:
\usepackage{subcaption} %% For complex figures with subfigures/subcaptions
\usepackage[
a4paper,
pdftex,  
pdfkeywords={},
pdfborder={0 0 0},
draft=false,
bookmarksnumbered,
bookmarks,
bookmarksdepth=2,
bookmarksopenlevel=2,
bookmarksopen]{hyperref}

\usepackage[normalem]{ulem}
\usepackage{stmaryrd}
\usepackage{tikz}
\usetikzlibrary{shapes,calc}
\usepackage{url}
\usepackage{multirow}
\usepackage{multicol}
\usepackage{rotating}
\usepackage{listings}
\usepackage[utf8]{inputenc}
\usepackage[T1]{fontenc}
\usepackage{bussproofs}
\usepackage{hyphenat}

\usepackage{enumitem} 
\usepackage{fancyvrb}
\usepackage{listings}

\usepackage{color}

\definecolor{light-gray}{gray}{0.85}

\include{myCommandsSubstSets}
\makeatletter

\EnableBpAbbreviations

\begin{document}

\title 
{
	%rensets 
	Rensets and Renaming-Based \\Recursion for Syntax with Bindings
	%A Substitution-Based Foundation \\for Syntax with Bindings
	%What is the World's Strongest Nominal Recursor?
	\vspace*{-1.8ex}
} 
%\subtitle{A Framework for Comparing Recursor Expressiveness} %Recursor 
%
\author{Andrei Popescu %\inst{1} 
		\vspace*{-0.7ex}
}
\institute{
	Department of Computer Science, 
	University of Sheffield, UK
}

\maketitle 

 \vspace*{-3ex}
\begin{abstract}
	I introduce \emph{renaming-enriched sets} (\emph{rensets} for short), which are algebraic structures axiomatizing 
	fundamental properties of renaming (also known as variable-for-variable substitution) on syntax with bindings. 
	%These structures are proposed as a foundation for syntax with bindings. 
	%
	Rensets compare favorably in some respects with the well-known foundation based on nominal sets. In particular, renaming is a more fundamental %(and more self-contained) 
	operator than the nominal swapping  
	operator and enjoys a simpler, equationally expressed relationship with the variable-freshness predicate. 
	Together with some natural axioms matching properties of the syntactic constructors, rensets yield a truly minimalistic characterization of $\lambda$-calculus terms as an abstract datatype---one involving an infinite set  
	of \emph{unconditional equations}, referring only to the most fundamental term operators: the constructors and renaming.   
	This characterization yields a recursion principle, which (similarly to the case of nominal sets) can be improved by incorporating Barendregt's variable convention. 
	When interpreting syntax in semantic domains, my renaming-based recursor is easier to deploy than the nominal recursor. 
   My results have been validated with the proof assistant Isabelle/HOL. 
\end{abstract}

%% Keywords
%% comma separated list
%\keywords{syntax with bindings, 
%	recursors, 
%	nominal logic, 
%	alpha-equivalence, 
%	formalization, 
%	Isabelle/HOL}  %% \keywords is optional

\section{Introduction}
\label{sec-intro}
\vspace*{-1ex}

Formal reasoning about syntax with bindings 
%(such as that of the $\lambda$-calculus or first-order logic
% \cite{bar-lam}
%) 
is necessary for the meta-theory of logics, %formal 
calculi and programming languages, and is 
notoriously error-prone. 
A great deal of research has been put into %convenient and powerful
 formal frameworks that make the specification of, and the reasoning 
 about bindings %of syntax with bindings 
 more manageable. 
%facilitate the specification of, and reasoning about syntax with bindings. %, the definition of functions and relations over this syntax, and the interpretation of syntax in semantic domains. 

%\subsection{%The 
%	Broader context: reasoning paradigms for syntax with bindings}
%	Nominal versus HOAS versus de Bruijn recursors}
%\label{subsec-paradigms}

%Today, 
Researchers wishing to formalize work involving syntax with 
bindings 
must choose a paradigm for representing and manipulating syntax---typically a variant of one of the ``big three'': 
nameful (sometimes called ``nominal'' reflecting its best known incarnation, nominal logic \cite{DBLP:conf/lics/GabbayP99,pitts-AlphaStructural}), nameless (De Bruijn) \cite{bru-lam,allais-bindingsByDependentTypes-agda,DBLP:conf/itp/SchaferTS15,DBLP:phd/dnb/Stark20} and higher-order abstract syntax (HOAS)  \cite{pfe-hoas,har-fra,pau-genTh,DBLP:journals/corr/FeltyMP15a,DBLP:journals/jar/FeltyMP15}.   
\leftOut{
The end results being formalized %themselves 
are (or at least should be) fundamentally the same regardless of the paradigm choice.
%---e.g., one proves using HOAS the Church-Rosser theorem for the $\lambda$-calculus, and not the "HOAS Church-Rosser theorem" for the "HOAS $\lambda$-calculus". 
However, this does not mean that these paradigms are equivalent %, or ``isomorphic'' 
from the proof developer's perspective. 
This is %largely 
because each paradigm offers 
different operators for manipulating syntax and different definitional and reasoning principles (e.g., for recursing or inducting over the structure of the syntax). %
} % end leftOut
Each paradigm has distinct advantages and drawbacks compared with each of the others, some discussed at length, e.g., in \cite{DBLP:journals/entcs/BerghoferU07}, \cite{DBLP:journals/jfp/AbelAHPMSS19} and \cite[\S8.5]{DBLP:journals/jar/GheriP20}. And there are also hybrid approaches, which combine some of the advantages   \cite{locallyNameless,DBLP:journals/jar/PollackSR12,momFelty-Hybrid4,pop-HOASOnFOAS}.  

A %major 
significant
advantage of the nameful paradigm is that it stays close to the way one informally defines and manipulates syntax when describing %formal 
systems in textbooks and research papers%  %and programming languages
---where the %(object-level) 
binding variables are explicitly indicated. This can in principle ensure transparency of the formalization and allows the formalizer to focus on the high-level ideas. % rather than on a specific encoding of syntax.  
%And indeed, the nameful approach, especially in the nominal logic incarnation, is very popular among formal proof developers---with the work on Isabelle/Nominal \cite{UrbanTasson} being honored by a CADE Skolem (test of time) award \cite{cadeSkolem}. 
%
%as witnessed by the large number of developments based on Nominal, acknowledged in 2015 by a CADE test-of-time award \cite{}. 
%by the original paper on Nominal having received the Skolem Award (the test-of-time award from the CADE) in 2015. 
%
However, %this 
it only works if 
the %main 
technical challenge faced by the nameful paradigm is properly addressed:
% at the level of conceptual design and tool support: 
enabling the seamless definition and manipulation of concepts ``up to alpha-equivalence'', i.e., in such a way that the names of the 
bound variables are (present but nevertheless) inconsequential. %, i.e., using terms rather than pre-terms. 
%can be summarized as  allowing to manipulate variables while maintaining well-definedness on terms (i.e., on pre-terms quotiented to alpha-equivalence), or, phrased another way, maintaining invariance under alpha-equivalence. 
This is particularly stringent in the case of recursion due to the binding constructors of terms not being free, hence not being \emph{a priori} traversable recursively---in that simply writing some recursive clauses that traverse the constructors is not \emph{a priori} guaranteed to produce a correct definition, but needs certain favorable conditions.  
The problem has been addressed by researchers in the form of tailored \emph{nameful recursors} \cite{DBLP:conf/lics/GabbayP99,pitts-AlphaStructural,urbanNominalRec,DBLP:journals/jar/Urban08,primrecFOAS-Norrish04,popescu-thesis}, which are theorems that identify such favorable conditions and, based on them,  guarantee %, under certain condition
the existence of functions that recurse over the non-free constructors. 
%that behave \emph{as if} defined recursively on %the 
%constructors. 
\looseness=-1 

In this paper, I make a contribution to the nameful paradigm in general, and to nameful recursion in particular. I introduce \emph{rensets}, which are algebraic structures axiomatizing the properties of renaming, also known as variable-for-variable substitution, on terms with bindings (Section~
\ref{sec-substSet}).  
Rensets differ from nominal sets (Section~\ref{subsec-nominal}), which form the foundation of nominal logic, by their focus on (not necessarily injective) renaming rather than swapping (or permutation).  
Similarly to nominal sets, rensets are %quite 
pervasive: Not only do the variables and terms form rensets, but so do any container-type combinations of rensets. % (Example~\ref{exa-containers-substSet}).  
\looseness=-1

While lacking the pleasant symmetry %specific to 
of swapping, my axiomatization of renaming has its %own 
advantages. First, renaming %emerges as 
is more fundamental than swapping because, at an abstract axiomatic level, renaming can define swapping but not 
vice versa %the other way around 
(Section~\ref{sec-substToNom}). 
The second advantage is about the ability to define another central operator: the variable freshness predicate. 
%that expressing the freshness of a variable for an object (such as a term). 
While the definability of freshness from swapping %has been 
is a signature trait of nominal logic, my renaming-based alternative fares even better: In rensets freshness 
%is not only definable, but it is definable 
has a simple, first-order definition (Section~\ref{sec-substSet}). This contrasts the nominal logic definition, which %necessarily 
involves a second-order statement about (co)finiteness of a set of variables.
%  (the discussion following Prop.~\ref{prop-freshIff}).   
%
The third advantage is largely a consequence of the second: Rensets enriched with constructor-like operators facilitate an %the first (to my knowledge) 
equational characterization of terms with bindings (using an infinite set of unconditional equations), which does not seem possible for swapping 
% is not available for swapping via nominal sets 
 (Section~\ref{subsec-eqChar}).  
This %in turn 
produces a recursion principle (Section~\ref{subsec-recPrinc}) which, like the nominal recursor, caters for Barendregt's variable convention, 
and in some cases is easier to apply than the nominal recursor---%namely
for example when interpreting syntax in semantic domains (Section~\ref{subsec-semInt}).  
\looseness=-1

%To my knowledge, this work provides the first purely equational characterization of terms with bindings as an ``ordinary'' datatype modulo equations that refer to the most fundamental operations only: the constructors and renaming. This 
%complements, and to some extent simplifies, 
%simplifies other characterizations 
%from the literature, which either abandon the simplicity of the underlying (standard) set theory or employ specification mechanisms that are more complex than equational theories (Section~\ref{sec-conclRelWork}).  

In summary, I argue that my renaming-based axiomatization 
%alternative to the swapping focus of nominal logic 
offers %a range of 
some benefits that %can 
strengthen the arsenal of the nameful paradigm: a simpler representation of freshness, a %new 
minimalistic equational characterization of terms, and a convenient recursion principle.  My results %have been 
are established with high %degree of 
confidence thanks to having been mechanized in Isabelle/HOL \cite{LNCS2283}. The mechanization is available \cite{pop-rensets-afp-2022} from 
Isabelle's Archive of Formal Proofs.  
% (\cite{substSets2022-form}, Appendix~\ref{app-isaForm}). 
%

Here is the structure of the rest of this paper: Section~\ref{sec-background} provides background on terms with bindings and on nominal logic.  Section~\ref{sec-substSet} introduces rensets and describes their basic properties. 
Section~\ref{sec-substToNom} establishes a formal connection 
%and discusses some points of comparison 
to nominal sets. 
Section~\ref{sec-recursion} discusses substitutive-set-based recursion. %Section~\ref{sec-isaForm} discusses my Isabelle formalization and sketches the main ideas behind the proofs of my results.
Section~\ref{sec-conclRelWork} discusses related work.  
The appendix includes more examples and results, details on nominal sets, and a discussion of my Isabelle formalization.

\section{Background} %: Binding-Aware Nameful Recursion}
\label{sec-background}
\vspace*{-1ex}

This section recalls the terms of $\lambda$-calculus 
and their basic operators 
(\S\ref{subsec-terms}), and aspects of nominal logic including nominal sets and nominal recursion (\S\ref{subsec-nominal}). 

\vspace*{-1ex}
\subsection{Terms with bindings}
\label{subsec-terms}
\vspace*{-0.5ex}

I work with the paradigmatic syntax of (untyped) $\lambda$-calculus. However, my results generalize routinely to syntaxes specified by arbitrary binding signatures such as the ones in \cite[\S 2]{fio-abs}, \cite{pitts-AlphaStructural}, \cite{nominalTwo} or \cite{DBLP:journals/pacmpl/BlanchetteGPT19}. 

Let $\Var$ be a countably infinite set of variables, ranged over by $x,y,z$ etc. 
%This will be fixed all throughout the paper. 
%
%
The set $\Trm$ of \emph{$\lambda$-terms} (or \emph{terms} for short), ranged over by $t,t_1,t_2$ etc., 
is defined by the grammar \hspace*{6ex}
$
t \;::=\; \Vr\;x  \;\mid\; \Ap\;t_1\;t_2  \;\mid\; \Lm\;x\;t
$
\\
with the proviso that terms are equated (identified) modulo alpha-equivalence (also known as naming equivalence). 
Thus, for example, if $x \not= z \not= y$ then $\Lm\;x\;(\Ap\;(\Vr\;x)\;(\Vr\;z))$ and 
$\Lm\;y\;(\Ap\;(\Vr\;y)\;(\Vr\;z))$ are considered to be the same term. 
I will often omit $\Vr$ when writing terms, as in, e.g., $\Lm\;x\;x$. 
%For example,, $\Lm\;x\;x$ is considered 
%	\emph{equal} to 
%	$\Lm\;y\;y$ %even if $x\not=y$

What the above specification means is (something equivalent to) the following: 
%
%\begin{myitem}
%	\item[(1)] 
One first defines the set $\PTrm$ of 
\emph{pre-terms} %(sometimes called ``raw terms'') 
as freely generated by the grammar %following grammar: 
$
p \;::=\; \PVr\;x  \;\mid\; \PAp\;p_1\;p_2  \;\mid\; \PLm\;x\;p
$. 
%
%\item[(2)] 
Then one defines the alpha-equivalence relation 
$\equiv\; : \PTrm \ra \PTrm \ra \Bool$ 
inductively, proves that it 
is an equivalence, and defines $\Trm$ by quotienting $\PTrm$ to alpha-equivalence, i.e., $\Trm = \PTrm/\!\equiv$. 
%
%\item[(3)] 
Finally, one proves that the pre-term constructors are compatible with $\equiv$, 
and defines the term counterpart of these constructors: 
$\Vr : \Var \ra \Trm$, $\Ap : \Trm \ra \Trm \ra \Trm$ 
and $\Lm : \Var \ra \Trm \ra \Trm$. 
%
%\end{myitem}
%(There are alternative definitions that do not go through alpha-equivalence, but in the end give the same notion of term.)

%In addition to the constructors, other standard operations and relations can be defined on terms, such as swapping, renaming and freshness; often, these are first defined on pre-terms, then lifted to terms (similarly to how constructors were lifted). I can prove a number of properties that are useful for manipulating terms, which do not mention pre-terms or alpha-equivalence---they just talk about terms and operations and relations on them, including term equality. 
%(Many of these properties will be recalled in Section~\ref{subsec-operPropsLam}.) Custom induction and recursion principles for terms are among these useful properties. 

The above constructions are %tedious
technical, but well-understood, and can be fully automated for an arbitrary syntax with bindings (not just that of  $\lambda$-calculus); and tools such as the Isabelle/Nominal package 
\cite{UrbanTasson,nominalTwo}  
provide this automation, hiding pre-terms completely from the end user. % 
%
%\footnote{There are alternative definitions which do not go through alpha-equivalence. For example, quasi-terms can be defined using functions from variables to model $\lambda$-abstraction, and then ``exotic'' quasi-terms are excluded to obtain terms, or using De Bruijn indices for bound variables and actual variables for the free variables only, etc. All these alternatives give us the 
%	same notion of term---see also the related work section.} 
% 
In formal and informal presentations alike, one usually prefers to forget about pre-terms, %completely, 
and work with terms only. 
%Working with terms rather than pre-terms
This has several advantages,
 including (1) being able to formalize concepts at the right abstraction level (since in most applications the naming of bound variables should be inconsequential) and (2) the renaming operator being well-behaved. 
%This is why most formal and informal developments prefer terms. 
%TODO: citations supporting these. 
%
However, there are some difficulties that need to be overcome when working with terms, and in this paper I focus on one of the major ones: 
%, I discuss some mechanisms that are particularly difficult to get 
%right for terms, namely 
providing recursion principles, i.e., mechanisms for defining 
functions by recursing over terms. This difficulty arises essentially because, unlike in the case of pre-term constructors, the binding constructor for terms is not free. 

%I will discuss and classify some %ingenious technical solutions to this problem stemming from the nominal logic tradition. 

%For the rest of this paper, I will focus on terms and mostly forget about pre-terms---the latter will only occasionally show up when I discuss ideas and intuitions. I will use ``recursor'' as a synonym for ``recursion principle''. 

The main characters of my paper will be (generalizations of)
some common operations and relations on 
$\Trm$, namely: % endowed with algebraic structure, given by operators such as:  
\begin{myitem}
	\item the constructors $\Vr : \Var \ra \Trm$, $\Ap : \Trm \ra \Trm \ra \Trm$ 
	and $\Lm : \Var \ra \Trm \ra \Trm$ 
	\item (capture-avoiding) renaming, also known as (capture-avoiding) substitution  of variables for variables
	$\_[\_/\!\_] : \Trm \ra \Var \ra \Var \ra \Trm$;   
	e.g., we have $(\Lm\;x\;(\Ap\;x\;y))\;[x / y] = \Lm\;x'\;(\Ap\;x'\;x)$
	\item swapping  
	$\sww : \Trm \ra \Var \ra \Var \ra \Trm$; 
	e.g., we have %\\
	\\$(\Lm\;x\;(\Ap\;x\;y))\,[x \sw y] = 
	\Lm\;y\;(\Ap\;y\;x)$
	\item the free-variable operator 
	$\FV : \Trm \ra \Pow(\Var)$ (where $\Pow(\Var)$ is the powerset of $\Var$); 
	e.g., %
	we have $\FV (\Lm\;x\;(\Ap\;y\;x)) = \{y\}$ 
	\item freshness $\_\fresh\_ : \Var \ra \Trm \ra \Bool$; 
	e.g., %
	we have $x \,\fresh\, (\Lm\;x\;x)$; and 
	assuming $x\not=y$, we have $\neg\;x \,\fresh\, (\Lm\;y\;x)$ 
\end{myitem}

The free-variable and freshness operators are of course related: %two faces of the same coin: 
A variable $x$ is fresh for a term $t$ (i.e., $x \,\fresh\, t$) 
if and only if it is not free in $t$ (i.e., $x \notin \FV(t)$). 
%
%Note that 
The renaming operator $\_[\_/\!\_] : \Trm \ra \Var \ra \Var \ra \Trm$ substitutes (in terms) \emph{variables} for variables, not terms for variables. (But an algebraization of term-for-variable substitution is discussed in Appendix~\ref{app-SSsets}.)

%In this paper, 
%I will not give the definitions of the above well-known operations and relations on terms, but count on the reader's familiarity with them and hope that the above examples %(and their subsequently listed properties in Section~\ref{subsec-operPropsLam}) 
%clarify any possible confusion; the actual definitions can be performed in several equivalent ways---see, e.g., \cite{bar-lam}, \cite{pitts-AlphaStructural}, \cite{popescu-thesis}[\S1.4.1].    

\vspace*{-1ex}
\subsection{Background on nominal logic}
\label{subsec-nominal}
\vspace*{-0.5ex}

%This paper's contribution is on the one hand inspired by nominal logic, and on the other hand offers an alternative to the nominal view. 
%So it is useful to recall some nominal logic concepts. 
I will employ a %broadly accessible (and broadly applicable) 
formulation of nominal logic %concepts due to Pitts
 \cite{pitts-AlphaStructural,urbanNominalRec,DBLP:journals/iandc/Pitts03} 
%whose work is based on previous work by Pitts and Gabbay and 
%by 
%and Urban et al.~\cite{urbanNominalRec},   
%todo: Some more basics of nominal logic? 
%which 
that does not require any special 
logical foundation, e.g., axiomatic nominal set theory. 
For simplicity, I prefer the swapping-based formulation %in the style of
\cite{DBLP:journals/iandc/Pitts03} 
to the equivalent permutation-based formulation---Appendix~\ref{app-swapVsPermNominal} gives details on these two alternatives. 

A \emph{pre-nominal set} is %a set equipped with a swapping action, i.e., 
a pair $\AA = (A,\sww)$ where $A$ is a set and 
$\sww : A \ra \Perm \ra A$ is a function called \emph{the swapping operator of $\AA$} satisfying the following properties for all $a\in A$ and $x,x_1,x_2,y_1,y_2\in\Var$: 
\begin{myitemmm} 
\item[Identity: \ ] $a[x\sw x] = a$
\item[Involution: \ ] $a[x_1\sw x_2][x_1\sw x_2] = a$
\item[Compositionality: \ ]
$a[x_1 \sw x_2] [y_1\sw y_2] = a[y_1\sw y_2][(x_1[y_1\sw y_2]) \sw (x_2[y_1\sw y_2] )]$ 
\end{myitemmm} 

Given a pre-nominal set $\AA = (A,\sww)$, an element $a\in A$ and a set $X\su\Var$, one says that \emph{$a$ is supported by $X$} 
%, or that \emph{$X$ supports $a$}, 
if 
$a [x \sw y] = a$ holds for all $x,y\in\Var$ such that $x,y\notin X$.  
An element $a\in A$ is called \emph{finitely supported} if there exists a finite set $X\su A$ such that $a$ is supported by $X$. 
A \emph{nominal set} is a pre-nominal set $\AA = (A,\sww)$ such that every element of $a$ is finitely supported.
If $\AA = (A,\sww)$ is a nominal set and $a\in A$, then the smallest set $X\su A$ such that $a$ is supported by $X$ exists, and is denoted by $\supp^\AA\,a$ and called the \emph{support of $a$}.  One calls a variable $x$ \emph{fresh for} $a$, written $x \,\fresh\,a$, if 
$x\notin \supp^\AA\,a$.  

An alternative, more direct definition of freshness (which is preferred, e.g., by Isabelle/Nominal \cite{UrbanTasson,nominalTwo}) is provided by the following 
proposition:

\begin{prop}\rm\label{prop-nomFreshIff}
	For any nominal set $\AA = (A,\sww)$ and any $x\in\Var$ and $a\in A$, it holds that $x \,\fresh\, a$ if and only if the set $\{y \mid a[y\sw x] \not= a\}$ is finite. 
\end{prop}

Given two pre-nominal sets $\AA = (A,\sww)$ and $\BB = (B,\sww)$,  the set $F = (A \ra B)$ of functions from $A$ to $B$  %naturally forms a 
becomes a pre-nominal set  $\FF = (F,\sww)$ by defining $f[x \sw y]$ %to be the function that sends 
to send each $a\in A$ to $(f(a[x\sw y]))[x \sw y]$. $\FF$ is not a nominal set because not all functions are finitely supported (though   
of course 
one obtains a nominal set by restricting to finitely supported functions). 

The set of terms together with their swapping operator, $(\Trm,\sww)$, forms a nominal set, where the support of a term is precisely its set of free variables. 
%The support of a term $t$ consists precisely of its free variables, i.e., the variables $x$ that are not fresh for $t$ (i.e., $\neg\; x \,\fresh\, t$).  
However, the power of nominal logic resides in the fact that not only the set of terms, but also many other sets can be organized as nominal sets---including the 
target domains of many functions one may wish to define on terms.  This gives rise to a convenient mechanism for defining functions recursively on terms: 
%Nominal logic provides induction and recursion principles that are binding-aware, i.e., tailored to handle syntax with bindings smoothly. 
%Here is the nominal recursion principle for the syntax of $\lambda$-calculus: 

%Part of the 

\begin{thm} \rm  \cite{pitts-AlphaStructural}
	\label{thm-pittsRec}
	Let $\AA = (A,\_[\_])$ be a nominal set and let $\Vr^\AA : \Var \ra A$, 
	$\Ap^\AA : A \ra A \ra A$ and 
	$\Lm^\AA : \Var \ra A \ra A$ be some functions, 
	all supported by a finite set $X$ of variables 
	and with $\Lm^\AA$ satisfying %such that 
	the following freshness condition for binders (FCB): %holds: 
	There exists $x\in\Var$ such that $x\notin X$ and $x \,\fresh\,\, \Lm^\AA\;x\;a$ for all $a \in A$. 

	Then there exists a unique function $f: \Trm \ra A$ that is supported by $X$ and such that the following hold for all $x\in\Var$ and $t_1,t_2,t\in \Trm$: 
	\begin{myitemm}
		\item[(i)]  $f\,(\Vr\;x) = \Vr^\AA\;x$
		%\item[(ii)]  
		\hspace*{20ex}
		(ii) $f\,(\Ap\;t_1\;t_2) = \Ap^\AA\,(f\;t_1)\,(f\;t_2)$
		\item[(iii)] $f\,(\Lm\;x\;t) = \Lm^\AA\;x\;(f\;t)$ if $x\notin X$
	\end{myitemm}
\end{thm}

A useful feature of %nominal-style binding-aware 
nominal recursion %(and also of nominal induction) 
is the support for Barendregt's famous \emph{variable convention} \cite{bar-lam}[p.~26]: 
``If [the terms] $t_1,\ldots,t_n$ occur in a certain mathematical context (e.g. definition, proof), then in these terms all bound variables are chosen to be different from the free variables.''
The above recursion principle adheres to this convention by fixing a finite set $X$ of variables meant to be free in the definition context and 
guaranteeing that the bound variables in the definitional clauses are distinct from them. Formally, the target domain operators $\Vr^\AA$, $\Ap^\AA$ and 
$\Lm^\AA$ are supported by $X$, and 
the clause for $\lambda$-abstraction is conditioned by the binding variable $x$ being outside of $X$.  
(The Barendregt convention is also present in 
nominal logic %as binding-aware
via induction principles \cite{pitts-AlphaStructural,UrbanTasson,nominalTwo,urban-Barendregt}.)   
%, for both structural induction \cite{pitts-AlphaStructural,UrbanTasson,nominalTwo} and rule induction \cite{urban-Barendregt}.

\section{Rensets}
\label{sec-substSet}
\vspace*{-1ex}

This section introduces rensets, 
an alternative to nominal sets that axiomatize renaming rather than swapping or permutation. 

A \emph{renaming-enriched set} (\emph{renset} for short) is a pair $\AA = (A,\_[\_/\!\_])$  where  
$A$ is a set and $\_[\_/\!\_] : A \ra \Var \ra \Var  \ra A$ is an operator 
%that satisfies the following properties:
%, assumed to be universally quantified 
such that the following hold for all   $x,x_1,x_2,\alb x_3,y,y_1,y_2 \in 
\Var$ and $a\in A$:%\footnote{Henceforth, I will tacitly make such universal quantification assumptions for the properties that I state.} 
\begin{myitemmmm}
	\item[Identity: \ ] $a[x/x] = a$
	\item[Idempotence: \ ] If $x_1\not=y$ then $a[x_1/y][x_2/y] = a[x_1/y]$
	\item[Chaining: \ ] If $y\not=x_2$ then $a[y/x_2][x_2/x_1] [x_3/x_2]= a[y/x_2][x_3/x_1]$
	\item[Commutativity: \ ]
	If $x_2 \not= y_1 \not= x_1 \not= y_2$ then 
	$a[x_2/x_1] [y_2/y_1]= a[y_2/y_1][x_2/x_1]$
\end{myitemmmm} 
Let us call $A$ the \emph{carrier} of $\AA$ and $\_[\_/\!\_] $ the \emph{renaming operator} of $\AA$.  Similarly to the case of terms, we think of the elements $a\in A$ as some kind of variable-bearing entities and of 
$a[y/x]$ as the result of substituting $x$ with $y$ in $a$.  
With this intuition, the above properties are natural:
Identity says that substituting a variable with itself has no effect. 
Idempotence acknowledges the fact that, after its renaming, a variable $y$ is no longer there, so substituting it again has no effect. 
Chaining says that a chain of renamings $x_3/x_2/x_1$ has the same effect as the end-to-end renaming $x_3/x_1$ provided there is no interference from $x_2$, which is ensured by initially substituting $x_2$ with some other variable $y$. 
Finally, Commutativity allows the reordering of any two independent renamings. 

\vspace*{1ex}
%\vspace*{-0.5ex}
\noindent
\textbf{Examples}  
%\begin{exa}\rm \label{exa-vars-substSet}
	 $(\Var,\_[\_/\!\_])$ and $(\Trm,\_[\_/\!\_])$, the sets of variables and %respectively 
	 terms with the standard renaming operator on them, form 
	 rensets.
%	 (where $x[y/z]$ is $y$ if $x=z$ and $x$ otherwise), forms a renset. 
%\end{exa}
%
%\begin{exa}\rm \label{exa-terms-substSet}
%$(\Trm,\_[\_/\!\_])$, the set of terms with the standard (capture-avoiding) renaming operator, forms a renset. 
%\end{exa}
%
%\begin{exa}\rm \label{exa-containers-substSet}
	Moreover, given any functor $F$ on the category of sets and %given 
	a renset 
	$\AA = (A,\_[\_/\!\_])$, let us define the renset 
	$F\,\AA = (F\,A,\_[\_/\!\_])$ as follows: 
	for any $k \in F\,A$ and $x,y\in\Var$, $k[x/y] = F\,(\_[x/y])\,k$, where the last occurrence of $F$ refers to the action of the functor on morphisms. 
%	Indeed, all the desired equations hold for $F\,\AA$ thanks to $F$'s functoriality and the fact that they hold for $\AA$. 
	%
	This means that one can freely build 
new rensets from existing ones using container types (which are particular kinds of functors)---e.g., lists, sets, trees etc. Another way to put it: Rensets are closed under datatype and codatatype constructions \cite{traytel-et-al-2012}.  
%\end{exa}

\vspace*{1ex}
In what follows, let us fix a renset 
$\AA = (A,\_[\_/\!\_])$. 
One can define the notion of freshness of a variable for an element of $a$ in %a style that is similar to 
the style of nominal logic. %, but using renaming instead of swapping. 
%Moreover, %as I show next 
But the next proposition shows that simpler formulations are available. 

\begin{prop}\rm \label{prop-freshIff}
	The following are equivalent:
	\\(1) The set $\{y \in \Var \mid a[y/x] \not= a\}$ is finite. 
	\\(2) $a[y/x] = a$ for all $y \in \Var$.
	\hspace*{7ex}
	(3) $a[y/x] = a$ for some $y \in \Var \sm \{x\}$. 
\end{prop}

Let us define the predicate $\_\,\fresh\,\!\_ : \Var \ra A \ra \Bool$ as follows: $x\,\fresh\, a$, read \emph{$x$ is fresh for $a$}, if either of Prop.~\ref{prop-freshIff}'s equivalent properties holds. 

Thus, points (1)--(3) above are three alternative formulations of  
$x\,\fresh\, a$, all referring to the lack of effect of substituting $y$ for $x$, expressed as $a[y/x] = a$: namely that this phenomenon affects  
(1) all but a finite number of variables $y$, 
(2) all variables $y$, or (3) some variable $y\not=x$. The first formulation is the most complex of the three---it is the nominal definition, but using renaming instead of swapping. 
The other two formulations do not have counterparts in nominal logic, essentially because swapping %does not have the same ability 
is not as %good 
``efficient'' as renaming %when it comes to 
at exposing freshness. 
%``making a variable be fresh''.  
%``turning a variable fresh''. 
In particular, (3) does not have a nominal counterpart because there is no single-swapping litmus test for freshness. 
The closest we can get to property (3) in a nominal set is the following: 
$x$ is fresh for $a$ if and only $a[y \sw x] = a$ holds for some fresh $y$---but this %uses 
needs freshness to explain freshness! 

\vspace*{1ex}
%\vspace*{-0.5ex}
\noindent
\textbf{Examples (continued)}  
For the rensets of variables and terms, freshness defined as above coincides with the expected operators: distinctness in the case of variables and standard freshness in the case of terms. 
And applying the definition of freshness to rensets obtained using finitary container types has 
similarly intuitive outcomes; for example, the freshness of a variable $x$ for a list of items  $[a_1,\ldots,a_n]$ means that $x$ is fresh for each item $a_i$ in the list.  

\vspace*{1ex}
Freshness satisfies some intuitive properties, which can be easily proved from its definition and the renset axioms. 
In particular, point (2) of the next proposition is the freshness-based version of the Chaining axiom.
%, which is perhaps a more familiar property from syntactic renaming. 

\begin{prop}\rm \label{prop-freshBasic}
	The following hold:
	\\(1) If $x\,\fresh\, a$ then $a[y/x] = a$ 
	\hspace*{6ex}
	(2) $x_2\, \fresh\, a$ then $a[x_2/x_1] [x_3/x_2]= a[x_3/x_1]$
	% fresh-based chaining
	\\(3) If $z\, \fresh\, a$ or $z=x$, and $x \,\fresh\, a$ or $z\not= y$, 
	then $z \,\fresh\, a[y/x]$ 
\end{prop}

\section{Connection %and Comparison 
	to Nominal Sets}
\label{sec-substToNom}
\vspace*{-1ex}

%All properties in Section~\ref{sec-substSet} follow from the 
So far I focused on consequences of the purely equational theory of rensets, without making any assumption about cardinality. But after additionally 
postulating a nominal-style finite support property, one can show that rensets give rise 
to nominal sets---which is what I will do in this section. 

%Perhaps familiar from nominal logic is the following alternative definition of freshness, whose analogue for rensets also holds. Given $a\in A$ and a set $X\su\Var$, let us say $a$ is \emph{supported} by $X$ if $\{y\in\Var \mid a[y/x] \not= x\} \su X$. 

%\begin{prop}\rm\label{prop-charFinSupp}
%	The following hold for any renset $\AA = (A,\_[\_])$ that satisfies Finite Support and any $a\in A$. 
%	\\(1) There exists the smallest set that supports $a$, denoted $\supp^\AA a$. 
%	\\(2) For all $x\in\Var$, it holds that $x \fresh\, a$ iff 
%	$x \notin \supp^\AA$. 
%\end{prop}

%\subsection{From renaming to swapping}
%\label{subsec-substToSwap}

Let us say that a renset $\AA = (A,\_[\_/\!\_])$ has the \emph{Finite Support} property if, for all $a\in A$, the set $\{x\in\Var \mid \neg\;x\,\fresh\, a\}$ is finite.  

Let $\AA = (A,\_[\_/\!\_])$ be a renset satisfying Finite Support. 
Let us define the swapping operator $\sww : A \ra \Var \ra \Var  \ra A$ as follows: 
$a[x_1\sw x_2] = a [y/x_1][x_1/x_2][x_2/y]$, where $y$ is a variable that is fresh for all the involved items, namely $y\notin\{x_1,x_2\}$ and $y \,\fresh\, a$. 
Indeed, this is how one would define swapping from renaming on terms: using a fresh auxiliary variable $y$, and exploiting that such a fresh $y$ exists and that its choice is immaterial for the end result. 
The next lemma shows that this style of definition also works abstractly, i.e., all it needs 
are 
the renset axioms plus Finite Support. 

\begin{mylemma}\rm \label{sw-choiceIrr}
	The following hold for all $x_1,x_2\in \Var$ and $a\in A$: 
\\(1) There exists $y\in \Var$ such that 
 $y\notin\{x_1,x_2\}$ and $y\, \fresh\, a$. 
 \\
 (2) For all $y,y' \in \Var$ such that  $y\notin\{x_1,x_2\}$, $y\, \fresh\, a$, 
 $y'\notin\{x_1,x_2\}$ and $y' \fresh\, a$, %we have that 
 $a [y/x_1][x_1/x_2][x_2/y] = a [y'/x_1][x_1/x_2][x_2/y']$. 
\end{mylemma}

And one indeed obtains an operator satisfying the nominal axioms: % of swapping: 
%Let us define $\_\sfresh\,\_$, the \emph{swapping-based freshness} operator, by taking 
%$x\,\sfresh\, a$ to mean that the set $\{y \mid a[y \sw x] \not= a\}$ is finite. 

\begin{prop}\rm \label{prop-sw-set-axioms}
If $(A,\_[\_/\!\_])$ is a renset satisfying Finite Support, then \linebreak
$(A,\sww)$ is a nominal set. Moreover, $(A,\_[\_/\!\_])$ and $(A,\sww)$ have the same notion of freshness, in that the freshness operator defined from renaming coincides with that defined from swapping. 
\end{prop}

%\subsection{Wrapping up the connections}
%\label{subsec-wrappingUpConnection}

The above construction is functorial, as I detail next. 
Given two nominal sets $\AA = (A,\sww)$ and $\BB = (B,\sww)$, a \emph{nominal morphism} $f:\AA \ra \BB$ is a function $f:A\ra B$ with the property that it commutes with swapping, in that $(f\;a)[x \sw y] = f(a[x \sw y])$ for all $a\in A$ and $x,y\in\Var$. Nominal sets and nominal morphisms form a category that I will denote by $\Nom$. 
Similarly, let us define 
a morphism $f:\AA \ra \BB$ between two rensets $\AA = (A,\_[\_/\!\_])$ and $\BB = (B,\_[\_])$ to be a function $f:A\ra B$ that commutes with renaming, yielding the category $\Sbs$ of rensets.  Let us write $\FSbs$ for the full subcategory of $\Sbs$ given by rensets that satisfy Finite Support. 
Let us define $F: \FSbs \ra \Nom$ to be an operator on objects and morphisms that sends each finite-support renset to the above described nominal set constructed from it, and sends each substitutive morphism to itself. 

\begin{thm} \rm \label{thm-connections}
	$F$ is a functor between $\FSbs$ and $\Nom$ which is injective on objects 
	and full and faithful (i.e., bijective on morphisms). 
\end{thm}

%An interesting question is 
One may ask whether it is also possible to make the trip back: from nominal %sets 
to rensets. The answer is negative, at least if %in case 
one wants to retain the same notion of freshness, i.e., have the freshness predicate defined in the nominal set be identical to the one defined in the resulting renset.  This is because swapping preserves the cardinality of the support, % of an element,
 whereas renaming must be allowed to change it 
 since it might perform a non-injective renaming. The following example captures this %intuitive 
idea:  

\vspace*{1ex}
\noindent
\textbf{Counterexample} 
%\begin{exa}\label{exa-noBackFromSwToSbs}\rm 
	Let $\AA = (A,\sww)$ be a nominal set such that all elements of $A$ have their support consisting of exactly two variables, $x$ and $y$ (with $x\not=y$). 
	(For example, $A$ can be the set of all terms with these free variables---this is indeed a nominal subset of the term nominal set because it is closed under 
	swapping.) 
	Assume for a contradiction that $\_[\_/\!\_]$ is an operation on $A$ 
	%(not necessarily equal to the standard renaming!) 
	that makes $(A,\_[\_/\!\_])$ a renset with its induced freshness operator equal to that of $\AA$. % (i.e., in this case, the standard freshness operator on terms).
	Then, by the definition of $A$, $a[y/x]$ needs to have exactly two non-fresh variables. But this is impossible, since by Prop.~\ref{prop-freshBasic}(3), all the variables different from $y$ (including $x$) must be fresh for $a[y/x]$.  
	In particular, $\AA$ is not in the image of the functor 
	$F:\FSbs\ra \Nom$, which is therefore not surjective on objects. 
%\end{exa}
\looseness=-1

\vspace*{1ex}
%\subsection{Summary: rensets versus nominal sets}
%\label{subsec-summary-SubstVsNominal}
%In summary, 
Thus, at an abstract algebraic level renaming can define swapping, but not the other way around. %---%\footnote{My statement refers to the possibility of an abstract definition within the axiomatizations of the two operators. On the other hand, of course for concrete terms both swapping and renaming are definable from the constructors by inductive or recursive mechanisms.} 
%which %means 
%reinforces the idea that renaming is the more fundamental operator 
%of the two. %than swapping. 
This is not too surprising, since swapping is fundamentally
 bijective whereas renaming is not; but it further validates
 % it is significant since 
 our axioms for renaming, highlighting their ability to define a well-behaved swapping. % operator. 

\section{Recursion Based on Rensets}
\label{sec-recursion}
\vspace*{-1ex}

Prop.~\ref{prop-freshIff} shows that, in rensets, 
renaming can define freshness using only equality and universal or existential quantification over variables---without needing any cardinality condition like in the case of swapping. 
As I am about to discuss, this forms the basis of a characterization of terms as the initial algebra of an equational theory (\S\ref{subsec-eqChar}) and an %easy to deploy
expressive  recursion principle (\S\ref{subsec-recPrinc}) that fares better than the nominal one for interpretations in semantic domains (\S\ref{subsec-semInt}).  

\vspace*{-1ex}
\subsection{Equational characterization of the term datatype}
%{Enriching rensets with compatible constructors}
\label{subsec-eqChar}
\vspace*{-0.5ex}

Rensets contain elements that are ``term-like''  in as much as there is a renaming operator on them satisfying %some 
familiar properties of renaming on terms. %\footnote{As discussed in Section~\ref{sec-substSet}'s examples, the set of variables also form such a ``term-like'' structure, in a trivial way; and so do any functors applied to term-like structures.} 
This similarity with terms can 
be %further 
strengthened by enriching rensets 
with operators having arities that match those of the term constructors. 

A \emph{constructor-enriched renset} (\emph{CE renset} for short) is a tuple $\AA = (A,\_[\_/\!\_],\alb\Vr^\AA,\Ap^\AA,\Lm^\AA)$ where:
\begin{myitem}
	\item  $(A,\_[\_/\!\_])$ is a renset
	\item  $\Vr^\AA : \Var \ra A$, 
	$\Ap^\AA : A \ra A \ra A$ and 
	$\Lm^\AA : \Var \ra A \ra A$ are functions 
\end{myitem}
such that the following hold for all $a,a_1,a_2\in A$ and $x,y,z\in\Var$:
\begin{myitemm}
	\item [(S1)] $(\Vr^\AA\;x)[y/z] = \Vr^\AA (x[y/z])$
	\hspace*{3ex}
	\item [(S2)]  $(\Ap^\AA\;a_1\;a_2)[y/z] = \Ap^\AA(a_1[y/z])\,(a_2[y/z])$
	\item [(S3)]  if $x\notin \{y,z\}$ then 
	$(\Lm^\AA\,x\;a)[y/z] = \Lm^\AA\,x\,(a[y/z])$
	\item [(S4)]  $(\Lm^\AA\,x\;a)[y/x] = \Lm^\AA\,x\;a$
	\item [(S5)]    if $z\not=y$ then $\Lm^\AA\,x\;(a[z/y]) = \Lm^\AA\,y\;(a[z/y][y/x])$
\end{myitemm}
Let us call $\Vr^\AA,\Ap^\AA,\Lm^\AA$ the \emph{constructors} of $\AA$. 
(S1)--(S3) express the constructors' commutation with renaming (with capture-avoidance provisions in the case of (S3)),  
%in the style of 
%(like with substitutive morphisms but with side conditions %accounting for bindings),  
(S4) the lack of effect of substituting 
for a bound variable, and (S5) 
the possibility to rename a bound variable 
without changing the abstracted item (where the inner renaming of $z\not= y$ for $y$ ensures the freshness of the ``new name'' $y$, hence its lack of interference with the other names in the  ``term-like'' entity where the renaming takes place).  
All these %are well-known 
%properties 
are well-known to hold for terms: 

%\ \\
%{\bf Example \ref{exa-terms-substSet} (continued)}  
%\begin{exa} \rm \label{exa-terms-substSet2}
\vspace*{1ex}
\noindent
\textbf{Example} 
Terms with renaming and the constructors, namely $ (\Trm,\_[\_/\!\_],\alb\Vr,\Ap,\Lm)$, form a CE renset which will be denoted by $\TTrm$. 
%\end{exa} 

\vspace*{1ex}
As it turns out, the CE renset axioms capture exactly the term structure $\TTrm$, via initiality. 
The notion of \emph{CE substitutive morphism} $f:\AA \ra \BB$ 
between two CE rensets $\AA = (A,\_[\_/\!\_],\Vr^\AA,\Ap^\AA,\Lm^\AA)$ and $\BB = (B,\_[\_/\!\_],\alb\Vr^\BB,\Ap^\BB,\Lm^\BB)$ is the expected one: 
a function $f:A\ra B$ that is a substitutive morphism and also 
commutes 
%not only with subsitution (like a 
%substitutive morphism) but also 
with the constructors. Let us write 
$\SbsCE$ for the category of CE rensets and morphisms.

\begin{thm} \rm \label{thm-termInit}
$\TTrm$ %= (\Trm,\_[\_/\!\_],\Vr,\Ap,\Lm)$  
is the initial CE renset, i.e., initial object in $\SbsCE$. 
\end{thm}
\emph{Proof idea.} Let $\AA = (A,\_[\_/\!\_],\Vr^\AA,\Ap^\AA,\Lm^\AA)$  
be a CE renset. Instead of directly going after a function $f: \Trm \ra A$, one first inductively defines 
a relation $R : \Trm \ra A \ra \Bool$, with inductive clauses reflecting the desired properties concerning the commutation with the constructors, e.g., 
$\frac{R\;t\;a}{R\;(\Lm\;x\;t)\;(\Lm^\AA\;x\;a)}$.  It suffices to prove that $R$ is total and functional and preserves renaming, since that allows one to define a constructor- and renaming-preserving function (a morphism) $f$ by taking $f\;t$ to be the unique $a$ with $R\;t\;a$. 

Proving that $R$ is total is easy by standard induction on terms. 
Proving the other two properties, namely functionality and preservation of renaming, is more elaborate and requires their simultaneous proof %of these two properties 
together with a third property: 
that $R$ preserves freshness. The simultaneous three-property proof follows by a form of ``substitutive induction'' on terms: Given a predicate $\phi : \Trm \ra \Bool$, to show $\forall t\in \Trm.\;\phi\;t$ it suffices to show the following: (1) $\forall x\in\Var.\;\phi\;(\Vr\;x)$, (2) $\forall t_1,t_2\in\Trm.\;\phi\;t_1 \,\&\, \phi\;t_2 \lra \phi\,(\Ap\;t_1\;t_2)$, and (3) $\forall x\in\Var,\,t\in\Trm.\;(\forall s\in\Trm.\; \SConnect\;t\;s \lra \phi\;s) \lra \phi\,(\Lm\;x\;t)$, where $\SConnect\;t\;s$ means that $t$ is connected to $s$ by a chain of renamings. 

Roughly speaking, $R$ turns out to be functional because the $\lambda$-abstraction operator on the ``term-like'' inhabitants of $A$ is, thanks to the axioms of CE renset, at least as non-injective as (i.e., identifies at least as many items as) the $\lambda$-abstraction operator on terms.  
\qed

\vspace*{1ex}
Theorem~\ref{thm-termInit} is the central result of this paper, from both practical and theoretical perspectives.  Practically, it enables a useful form of recursion on terms (as I will discuss in the following sections).   
Theoretically, this is a characterization of terms as the initial algebra of an equational theory that only the most fundamental term operations, namely the constructors and renaming. The equational theory consists of the axioms of CE rensets (i.e., those of rensets plus (S1)--(S5)), which are an infinite set of unconditional equations---for example, axiom (S5) gives one equation for each pair of distinct variables $y,z$.  

It is instructive to compare this characterization with the one offered by nominal logic, namely by Theorem \ref{thm-pittsRec}. To do this, one first needs a lemma: % (which seems to be folklore in nominal logic): 

\begin{mylemma}\rm 
	\label{lem-funSupp}
	Let $f: A \ra B$ be a function between two nominal sets 
	$\AA = (A,\sww)$ and $\BB = (B,\sww)$ and $X$ a set of variables. Then $f$ is supported by $X$ if and only if $f(a[x\sw y]) = (f\,a)[x\sw y]$ for all $x,y\in \Var \sm X$. 
	%$\supp\,\sigma \cap X = \emptyset$. 
	%	\end{myitem} 
%	In particular, $f$ is supported by $\emptyset$ iff $f$ commutes with %(two-variable) 
%	swapping, 
%	i.e. %(in nominal logic terminology) $f$ 
%	is equivariant. 
\end{mylemma} 

Now Theorem \ref{thm-pittsRec} (with the variable avoidance set $X$ taken to be $\emptyset$) can be rephrased as an initiality statement, as I describe below. 

Let us define a \emph{constructor-enriched nominal set} (\emph{CE nominal set}) to be any  tuple $\AA = (A,\sww,\Vr^\AA,\Ap^\AA,\Lm^\AA)$ where
%\begin{myitem}
%	\item  
	$(A,\sww)$ is a nominal set and 
%	\item  
	$\Vr^\AA : \Var \ra A$, 
	$\Ap^\AA : A \ra A \ra A$, 
	$\Lm^\AA : \Var \ra A \ra A$ are operators on $A$ 
	%with arities matching those of the term constructors
%\end{myitem}
such that the following properties hold for all $a,a_1,a_2\in A$ and $x,y,z\in\Var$:
\begin{myitemm}
	\item [(N1)] $(\Vr^\AA\;x)[y \sw z] = \Vr^\AA (x[y \sw z])$
	\item [(N2)] $(\Ap^\AA\;a_1\;a_2)[y \sw z]= \Ap^\AA(a_1[y \sw z])\,(a_2[y \sw z])$
	\item [(N3)] $(\Lm^\AA\,x\;a)[y \sw z] = \Lm^\AA\;(x[y \sw z])\;(a[y \sw z])$
	\item [(N4)] $x \,\fresh\, \Lm\;x\;a$, i.e., 
	$\{y \in \Var \mid (\Lm\;x\;a)[y \sw x] \not= \Lm\;x\;a\}$ is finite. 
%	\item [\mbox{\normalfont{(N5)}}] Every element of $a$ is finitely supported, i.e., $\{x \in \Var \mid \neg\;x \fresh a\}$ is finite. 
\end{myitemm}

The notion of \emph{CE nominal morphism} is defined 
as the expected extension of that of nominal morphism: 
a function that commutes with swapping and the constructors.  
Let $\NomCE$ be the category of CE nominal sets %and CE nominal 
morphisms. %Now here is the rephrasing of nominal recursion as initiality:  

\begin{thm} \rm \label{thm-nominalnit} (\cite{pitts-AlphaStructural}, rephrased) 
	$(\Trm,\sww,\Vr,\Ap,\Lm)$ 
	is the initial CE \linebreak nominal set, i.e., the initial object in $\NomCE$. 
\end{thm}

The above theorem indeed corresponds exactly to Theorem \ref{thm-pittsRec} with $X=\emptyset$:
\begin{myitem}
	\item the conditions (N1)--(N3) in the definition of CE nominal sets correspond 
	(via Lemma~\ref{lem-funSupp}) to %the notion of 
	the constructors being supported by $\emptyset$ 
	\item (N4) is the freshness condition for binders %(with $X=\emptyset$)
	\item initiality, i.e., the existence of a unique morphism, is the same as the existence of the unique function $f: \Trm \ra A$ stipulated in Theorem~\ref{thm-pittsRec}: commutation with the constructors is %the same as 
	the Theorem~\ref{thm-pittsRec} conditions (i)--(iii), and commutation with swapping means 
	(via Lemma~\ref{lem-funSupp}) $f$ being supported by $\emptyset$. 
\end{myitem}

Unlike the renaming-based characterization of terms 
(Theorem~\ref{thm-termInit}), the nominal logic %initiality 
characterization (Theorem~\ref{thm-nominalnit}) is not purely equational. This is due to a combination of two factors: (1) two of the axioms
((N4) and the Finite Support condition) % from the definition of nominal set) 
referring to freshness and 
(2) the impossibility of expressing freshness equationally from swapping. %\footnote{Incidentally, the finite support property can actually be removed from the definition of CE nominal set, and Theorem~\ref{thm-nominalnit} still holds; but (N4) cannot be removed.} 
%The problem seems fundamental; I conjecture that the structure of terms with the constructors and swapping 
%cannot have a recursively enumerable equational axiomatization without using additional operation or relation symbols.  
The problem seems fundamental, in that the nominal characterization does not seem expressible purely equationally. 
By contrast, while the freshness idea is implicit in the CE renset axioms, the freshness predicate itself %is completely eliminated 
is absent from Theorem~\ref{thm-termInit}.   
%the initiality end result. 

\vspace*{-1ex}
\subsection{Barendregt-enhanced recursion principle}
\label{subsec-recPrinc}
\vspace*{-0.5ex}

While Theorem~\ref{thm-termInit} already gives a %fairly expressive 
recursion principle, it is possible to improve it by incorporating Barendregt's variable convention (in the style of Theorem~\ref{thm-pittsRec}): %\footnote{In addition to the Barendregt enhancement, it is also possible to extend the principle from iteration to full-fledged primitive recursion---this completely standard type of enhancement is discussed in Appendix~\ref{app-fullFledgedRec}.}
%, obtaining a renaming-based analogue of Theorem~\ref{thm-pittsRec}. 

\begin{thm} \rm  
	\label{thm-substRec}
	Let $X$ be a finite set, 
	$(A,\_[\_/\!\_])$ a renset and $\Vr^\AA : \Var \ra A$, 
	$\Ap^\AA : A \ra A \ra A$ and 
	$\Lm^\AA : \Var \ra A \ra A$ some functions 
	that satisfy the clauses (S1)--(S5) from the definition of CE renset, %, %but relativized to $X$:
		but only under the assumption that $x,y,z\notin X$. 
	%	 the variables making up the renaming,
	%	\\ TODO: not clear here
	%	\\
	%	 $y$ and $z$, are not in $X$---
	%
	%
	Then there exists a unique function $f: \Trm \ra A$
	such that th following hold: 
	\begin{myitemm}
		\item[(i)] $f\,(\Vr\;x) = \Vr^\AA\;x$
		\hspace*{18ex}
	%	\item[(ii)] 
	(ii) 
		$f\,(\Ap\;t_1\;t_2) = \Ap^\AA\,(f\;t_1)\,(f\;t_2)$
		\item[(iii)] $f\,(\Lm\;x\;t) = \Lm^\AA\;x\;(f\;t)$ if $x\notin X$
			\hspace*{0ex}
	%\item[(iv)] 
	(iv) $f\,(t[y/z]) = (f\;t)[y/z]$ if $y,z\notin X$
	\end{myitemm}
\end{thm}
\emph{Proof idea.} The constructions in the proof of Theorem~\ref{thm-termInit} can be adapted to avoid clashing with the finite set of variables $X$. For example, the clause for $\lambda$-abstraction 
in the inductive definition of the relation $R$  
becomes $\frac{x \not\in X\;\;\;\;\;\;\;\;R\;t\;a}{R\;(\Lm\;x\;t)\;(\Lm^\AA\,x\;a)}$ and preservation of  renaming and freshness are also formulated to avoid $X$. Totality is still ensured thanks to the possibility of renaming bound variables---in terms and inhabitants of $A$ alike (via the modified axiom (S5)).  \qed 

\vspace*{1ex}
The above theorem says that if the structure $\AA$ % = (A,\_[\_/\!\_],\Vr^\AA,\Ap^\AA,\Lm^\AA)$ 
is assumed to be ``almost'' a CE set, save for additional restrictions involving the avoidance of $X$,  
then there exists a unique ``almost''-morphism---satisfying the CE substitutive morphism conditions restricted so that the bound and renaming-participating variables avoid $X$. 
It is the renaming-based counterpart of the nominal Theorem~\ref{thm-pittsRec}. 

In regards to the relative expressiveness of these two recursion principles (Theorems~\ref{thm-substRec} and \ref{thm-pittsRec}), it seems difficult to find %a practical 
an example that is definable by one but not by the other. In particular, my principle 
can seamlessly define standard nominal examples 
 \cite{pitts-AlphaStructural,pitts_2013} such as the length of a term, the counting of $\lambda$-abstractions or of the free-variables occurrences, and term-for-variable
substitution---Appendix~\ref{app-moreRecExa} gives details.  
However, as I am about to discuss, I found an important class of examples where my renaming-based principle is significantly easier to deploy: that of interpreting syntax in semantic domains. % I discuss this next. 

\vspace*{-1ex}
\subsection{Extended example: semantic interpretation}
\label{subsec-semInt}
\vspace*{-0.5ex}

Semantic interpretations, also known as denotations (or denotational semantics), are pervasive in the meta-theory of logics and $\lambda$-calculi, for example when interpretating first-order logic (FOL) formulas in FOL models, or untyped or simply-typed $\lambda$-calculus or higher-order logic terms in specific models (such as full-frame or Henkin models).  In what follows, I will focus on $\lambda$-terms and Henkin models, but the ideas discussed apply broadly to any kind of statically scoped interpretation of terms or formulas involving binders. 

Let $D$ be a set and $\ap : D \ra D \ra D$ 
and $\lm: (D \ra D) \ra D$ be operators modeling semantic notions of application and abstraction.  %(just like in Example~\ref{exa-denot}), 
An environment will be a function $\xi: \Var \ra D$. 
%, whose purpose is to bind variables to elements of $D$. 
Given $x,y\in\Var$ and $d,e\in D$, let us write $\xi\<x:=d\>$ for $\xi$ updated with value $d$ for $x$ (i.e., acting like $\xi$ on all variables except for $x$ where it returns $d$); and let us write $\xi\<x:=d,y:=e\>$ instead of $\xi\<x:=d\>\<y:=e\>$. 

Say one wants to interpret terms in the semantic domain $D$ in the context of environments, i.e., 
define 
the %semantic interpretation 
function $\sem : \Trm \ra (\Var \ra D) \ra D$ that maps syntactic to semantic constructs; e.g., one would like to have:
\begin{myitem} 
\item $\sem\,(\Lm\;x\;(\Ap\;x\;x))\;\xi = \lm(d \mapsto \ap\;d\;d)$ (regardless of $\xi$)
\item $\sem\,(\Lm\;x\;(\Ap\;x\;y))\;\xi = \lm(d \mapsto \ap\;d\;(\xi\;y))$ (assuming $x\not=y$)
\end{myitem} 
where I use $d\mapsto\ldots$ to describe functions in $D\ra D$, e.g., 
$d \mapsto \ap\;d\;d$ is the function sending every $d\in D$ to 
$\ap\;d\;d$. 

The definition should therefore naturally go recursively by the %following 
clauses:
\begin{myitem}
	\item[(1)] $\sem\,(\Vr\;x)\,\xi = \xi\;x$
	%\item[(2)]  
	\hspace*{12ex}
	(2) $\sem\,(\Ap\;t_1\,t_2)\,\xi = \ap\,(\sem\;t_1\,\xi)\,(\sem\;t_2\,\xi)$
	\item[(3)]  $\sem\,(\Lm\;x\;t)\,\xi = \lm\,(d \mapsto \sem\;t\,(\xi\<x:= d\>))$
\end{myitem}
%
%Thus, a variable term $\Vr\;x$ would be interpreted as the value of $x$ in the current environment $\xi$, an application term would be interpreted using $\ap$, and an abstraction term $\Lm\;x\;t$ would be interpreted via $\lm$ applied to the function sending each element $d$ in the domain to the interpretation of $t$ with $x$ bound to $d$ in the environment---in short, the usual pattern of semantic interpretations.  

Of course, since $\Trm$ is not a free datatype, these clauses do not work out of the box, i.e., do not form a definition (yet)---this is where binding-aware recursion principles such as Theorems~\ref{thm-substRec} and \ref{thm-pittsRec} could step in. I will next try them both. 

The three clauses above already determine constructor operations $\Vr^\SS$, $\Ap^\SS$ and $\Lm^\SS$ on the set of interpretations, $I = (\Var \ra D) \ra D$, namely:
\begin{myitem}
	\item $\Vr^\SS : \Var \ra I$ by $\Vr^\SS\, x\;i\;\xi = \xi\;x$
	\item $\Ap^\SS : I \ra I \ra I$ by $\Ap^\SS\, i_1\,i_2\;\xi = \ap\,(i_1\,\xi)\,(i_2\,\xi)$
	\item  $\Lm^\SS :\Var \ra I \ra I$ by $\Lm^\SS\, x\;i\;\xi = \lm\,(d \mapsto i\,(\xi\<x := d\>))$  
\end{myitem} 

To apply the renaming-based recursion principle from Theorem~\ref{thm-substRec}, one must further define a renaming operator %$[\_/\!\_]^\SS$ 
on $I$. 
Since the only chance to successfully apply this principle is if $\sem$ commutes with renaming, the definition should be inspired by the question: How can $\sem(t[y/x])$ be determined from $\sem\;t$, $y$ and $x$? The answer is %It is not hard to see the answer---we need %to have
%\begin{myitem}
%	\item[(4)]  
	(4) $\sem\,(t[y / x])\,\xi = (\sem\;t)\;(\xi\<x:= \xi\;y\>)$, 
%\end{myitem}
yielding an operator $[\_/\!\_]^\SS : I \ra \Var \ra \Var \ra I$ defined by 
$i\,[y/x]^\SS\,\xi = i\,(\xi\<x:= \xi\;y\>)$. 
%---which is exactly the renaming operator in Example \ref{exa-denot}. 

It is not difficult to verify that $\SS= (I,[\_/\!\_]^\SS,\Vr^\SS,\Ap^\SS,\Lm^\SS)$ is a CE renset---for example, Isabelle's automatic methods discharge all the goals. 
%Sledgehammer proves automatically all the necessary equations. 
%
This means Theorem~\ref{thm-substRec} (or, since here one doesn't need 
%to avoid any variables, 
Barendregt's variable convention, 
already Theorem~\ref{thm-termInit}) is applicable, and gives us a unique function $\sem$ that commutes with the constructors, i.e., satisfies clauses (1)--(3) (which are instances of the clauses (i)--(iii) from Theorem~\ref{thm-substRec}), and additionally commutes with renaming, i.e., satisfies clause (4) (which is an instances of the clause (iv) from Theorem~\ref{thm-substRec}).  

\vspace*{0.5ex}
On the other hand, %if one wishes 
to apply nominal recursion for defining $\sem$, one must identify a swapping operator on $I$. Similarly to the case of renaming, this identification process is guided by the goal of determining $\sem(t[x \sw y])$ from $\sem\;t$, $x$ and $y$, leading to
%\begin{myitem}
%	\item[(4')]  
	(4') $\sem\,(t[x \sw y])\,\xi = \sem\;t\;(\xi\<x:=\xi\;y,y:=\xi\;x\>)$, 
%\end{myitem}
%
which yields the definition of $[\_\sw\_]^\SS$ by $i\,[x \sw y]^\SS\,\xi = i\,(\xi\<x:=\xi\;y,y:=\xi\;x\>)$. 
However, as pointed out by Pitts \cite[\S 6.3]{pitts-AlphaStructural} (in the slightly different context of interpreting simply-typed $\lambda$-calculus), the nominal recursor 
(Theorem~\ref{thm-pittsRec})  
does \emph{not} directly apply (hence neither does my reformulation based on CE nominal sets, Theorem~\ref{thm-nominalnit}). 
This is because, in my terminology, the structure $\SS= (I,[\_\sw\_]^\SS,\Vr^\SS,\Ap^\SS,\Lm^\SS)$ 
is not a CE nominal set. The problematic condition is  FCB (the freshness condition for binders), 
requiring that $x\;\fresh^\SS\, (\Lm^\SS\,x\;i)$ holds for all $i\in I$. 
Expanding the definition of $\,\fresh^\SS$ (the nominal definition of freshness from swapping, recalled in Section~\ref{subsec-nominal}) and the definitions of $[\_\sw\_]^\SS$ and $\Lm^\SS$, one can see that $x\;\fresh^\SS\, (\Lm^\SS\,x\;i)$ means the following: 
\\$\lm\;(d \mapsto i\,(\xi\< x:=\xi\,y,y:=\xi\,x\>\< x:=d\>)) = \lm\;(d \mapsto i\,(\xi\< x:= d\>))$,  
i.e., 
$\lm\;(d \mapsto i\,(\xi \< x:=d,y:=\xi\,x \>) = \lm\;(d \mapsto i\,(\xi\<x:= d\>))$,  
holds for all but a finite number of variables $y$. 

The only chance for the above to be true is if $i$, when applied to an environment, ignores the value of $y$ in that environment for all but a finite number of variables $y$; 
in other words, $i$ 
only analyzes the value of a finite number of variables in that environment---but this is not guaranteed to hold for arbitrary elements $i\in I$.  
%
%So whereas the elements in the image of $\sem$ would satisfy this property, $\sem$ is not yet defined! 
To repair this, Pitts engages in a form of induction-recursion \cite{DBLP:journals/jsyml/Dybjer00}, carving out from $I$ a smaller domain that is still large enough to interpret all terms, then proving that both FCB and the other axioms hold for this restricted domain. It all works out in the end, but the technicalities are quite involved. %---one may say too involved for a ``mere'' recursive definition. %not straightforward. 

Although FCB is not required by the renaming-based principle, note incidentally that this condition would actually be true (and immediate to check) if working with freshness %as
 defined not from swapping but from renaming. Indeed, the renaming-based version of $x \;\fresh^\SS\, (\Lm^\SS\,x\;i)$ says that
$\lm\;(d \mapsto i\,(\xi\<x:=\xi\,y\>\<x:=d\>)) = \lm\;(d \mapsto i\,(\xi\<x:= d\>))$ 
%
%$$\lm\;(d \mapsto i\,(\xi\<x:=d\>)) = \lm\;(d \mapsto i\,(\xi\<x:= d\>)) \;,$$ 
holds for all $y$ (or at least for some $y\not=x$)---which is immediate since $\xi\<x:=\xi\,y\>\<x:=d\> = \xi\<x:=d\>$. 
This further illustrates the idea that semantic domains %seem to 
`favor' renaming over swapping. 

% illustrates the convenience of the purely equational characterization of freshness in renaming algebras.  

In conclusion, for interpreting syntax in semantic domains,  
my renaming-based recursor is trivial to apply, whereas the nominal recursor requires some fairly involved %workarounds in terms of
 additional definitions and proofs.

\section{Conclusion and Related Work}
\label{sec-conclRelWork}
\vspace*{-1ex}

\renewcommand\tt[2]{\begin{tabular}{c}#1\\#2\end{tabular}}
\newcommand\ttt[3]{\begin{tabular}{c}#1\\#2\\#3\end{tabular}}
\newcommand\tttt[4]{\begin{tabular}{c}#1\\#2\\#3\\#4\end{tabular}}

%There is a vast literature of algebraic 

This paper introduced and studied rensets, contributing 
% leading to a twofold
%main 
%contribution: 
(1) theoretically, a minimalistic equational characterization of the datatype of terms with bindings and (2) practically, an addition to the %current 
formal arsenal for manipulating syntax with bindings.   
It is part of a longstanding line of work by myself and collaborators on 
%discovering
exploring convenient 
%line of work that investigates 
%sprung out of my and my collaborator's %(Blanchette and Traytel) 
%investigations of ever more 
%convenient %and expressive  
definition and 
reasoning principles for bindings %, usually formalized in Isabelle %---in a line of work
 \cite{popescu-thesis,gun-proper,pop-HOASOnFOAS,DBLP:conf/icfp/PopescuG11,DBLP:journals/jar/GheriP20},  
 %that spans 
%both authors' PhD theses \cite{popescu-thesis,gheri-thesis}. 
%Many of these ideas are being 
and will be incorporated into the ongoing implementation of a new Isabelle definitional package for binding-aware datatypes \cite{DBLP:journals/pacmpl/BlanchetteGPT19}.

%\subsection
\vspace*{1ex}
\noindent
\textbf{Initial model characterizations of the terms %with bindings 
	datatype.}
My results % improve on previous work %, both by ourselves and by others, 
%by providing  
provide a truly elementary characterization of terms with bindings, as an ``ordinary'' datatype specified by the fundamental operations only (the constructors plus variable-for-variable renaming) and some equations (those defining CE rensets). As far as specification simplicity goes, this is ``the next best thing'' after a completely free datatype such as those of natural numbers or lists. %, and yields an easy-to-apply recursion principle. 

Fig.~\ref{fig-relatedWork} shows previous characterizations from the literature, in which terms with bindings are identified as an initial model (or algebra) of some kind. 
For each of these, I indicate (1) the employed reasoning paradigm, (2) whether the initiality/recursion theorem features an extension with Barendregt's variable convention, (3) the underlying category (from where the carriers of the models are taken), (4) the operations and relations on terms to which the models must provide counterparts and (5) the properties required on the models. 

While some of these results enjoy elegant mathematical properties of intrinsic value, my main interest is in the recursors they enable, specifically in the %convenience and 
ease of deploying these recursors. %In other words,
That is, I am interested in how %difficult 
easy it is in principle 
to organize the target domain as a model of the requested type, hence obtain the desired morphism, i.e., get the recursive definition done.  
By this measure, elementary approaches relying on standard FOL-like  
models whose carriers are sets rather than 
pre-sheaves 
have an advantage. %, 
%with set carriers have an advantage over those based on pre-sheaf models, 
%and are easier to %implement 
%use in proof assistants. 
%
Also, it seems intuitive that a recursor is easier to apply 
if there are fewer operators, and fewer and structurally 
simpler properties required on its %associated 
models---although empirical evidence of successfully deploying the recursor in practice should complement the simplicity assessment, 
to ensure that simplicity is not sponsored by 
%accompanied by 
lack of expressiveness. 

The first column in Fig.~\ref{fig-relatedWork}'s table contains an influential representative of the nameless paradigm: the result obtained independently by Fiore 
et al.\ \cite{fio-abs} and Hofmann \cite{DBLP:conf/lics/Hofmann99} characterizing terms as initial in the category of algebras over the pre-sheaf topos $\Set^{\mathbb{F}}$, where ${\mathbb{F}}$ is the category of finite ordinals and %injections 
functions between them. % (a skeleton of the category of finite sets).
The operators required by %Fiore et al.'s 
algebras are the constructors, as well as 
the free-variable operator (implicitly as part of the separation on levels) and 
the injective renamings (as part of the functorial structure). 
The algebra's carrier is required to be a functor and 
the constructors to be natural transformations. 
There are several %simultaneous or subsequent 
variations of this approach, e.g., \cite{BirdP-lambda,DBLP:conf/csl/AltenkirchR99,DBLP:conf/lics/Hofmann99}, 
some implemented in proof assistants, e.g., \cite{allais-bindingsByDependentTypes-agda,allais-icfp2018,DBLP:conf/cpp/KaiserSS18}.
%, of Fiore et al.'s approach. 

\begin{figure}[t]
	\hspace*{-2.5ex}
	\footnotesize
	%	\centering
	%\hspace*{-5ex}
	\begin{tabular}{c|c|c|c|c|c|c|c}
		& \tttt{Fiore}{et al.\ \cite{fio-abs}}{Hofmann}{\cite{DBLP:conf/lics/Hofmann99}}
		& \tt{Pitts}{\cite{pitts-AlphaStructural}}
		& 
		\ttt{Urban}{et al.}{\cite{urbanNominalRec,DBLP:journals/jar/Urban08}}
		& 
		\tt{Norrish}{\cite{primrecFOAS-Norrish04}}
		& \ttt{Popescu}{$\&$Gunter}{\cite{DBLP:conf/icfp/PopescuG11}}
		& \ttt{Gheri$\&$}{Popescu}{\cite{DBLP:journals/jar/GheriP20}}
		& \tt{This}{paper}
		\\\hline
		\begin{tabular}{c}
			Paradigm 
		\end{tabular} 
		& nameless & nameful & nameful & nameful 
		& nameful & nameful & nameful 
		\\\hline
		\begin{tabular}{c}
			Barendregt? 
		\end{tabular} 
		& n/a & yes & yes & yes
		& no & no & yes
		\\\hline
		%underlying 
		\begin{tabular}{c}
			Underlying \\
			category 
		\end{tabular} 
		& $\Set^{\mathbb{F}}$ & $\Set$ & $\Set$  & $\Set$ 
		& $\Set$  & $\Set$  & $\Set$ 
		\\\hline
		%underlying 
		\begin{tabular}{c}
			Required \\
			operations/ \\
			relations 
		\end{tabular} 
		&  \begin{tabular}{c}
			ctors, \\
			rename, \\
			free-vars
		\end{tabular} 
		& \begin{tabular}{c}
			ctors, \\
			perm
		\end{tabular}
		& \begin{tabular}{c}
			ctors, \\
			perm
		\end{tabular}
		& \begin{tabular}{c}
			ctors, \\
			swap, \\
			free-vars
		\end{tabular}
		& \begin{tabular}{c}
			ctors, \\
			term/var 
			\\subst, \\
			fresh
		\end{tabular}
		& \begin{tabular}{c}
			ctors, \\
			swap, \\
			fresh
		\end{tabular}
		& \begin{tabular}{c}
			ctors, \\
			rename
		\end{tabular}
		%%%
		%%%
		\\\hline
		%underlying 
		\begin{tabular}{c}
			Required \\
			properties
		\end{tabular} 
		&  \begin{tabular}{c}
			functori-
			\\ality,\\
			naturality
		\end{tabular} 
		& \begin{tabular}{c}
			%(cond)\\
			Horn\\
			clauses,  
			\\
			fresh-def, \\
			fin-supp 
		\end{tabular}
		& \begin{tabular}{c}
			%	(cond)\\
			Horn\\
			clauses,  
			\\
			fresh-def 
		\end{tabular}
		& \begin{tabular}{c}
			%	(cond)\\
			Horn\\
			clauses 
		\end{tabular}
		&  \begin{tabular}{c}
			%	(cond)\\
			Horn\\
			clauses 
		\end{tabular}
		&  \begin{tabular}{c}
			%	(cond)\\
			Horn\\
			clauses 
		\end{tabular}
		& \begin{tabular}{c}
			equations
		\end{tabular}
	\end{tabular}
	\vspace*{-1.1ex}
	\caption{Initial %algebra/ 
		model characterizations of the datatype of terms with bindings
		\\{\footnotesize 
			``ctors'' = ``constructors'', 
			``perm'' = ``permutation'', 
			%``swap'' = ``swapping'', 
			``fresh'' = ``the freshness predicate'', 
			``fresh-def" = ``clause for defining the freshness predicate'', 
			``fin-supp'' = ``Finite Support''
			%	``cond'' = ``conditional'' 
		}
	}
	\label{fig-relatedWork}
	\vspace*{-7.65ex}
\end{figure}

The other columns refer to initiality results that are more closely related to mine. They take place within the nameful paradigm, and they all rely on elementary models (with set carriers). Pitts's already discussed nominal recursor \cite{pitts-AlphaStructural} (based on previous work by Gabbay and Pitts  \cite{DBLP:conf/lics/GabbayP99}) employs the constructors and permutation (or swapping), and requires that its models satisfy some Horn clauses for constructors, permutation and freshness, together with the second-order properties that (1) define freshness from swapping and (2) express Finite Support. Urban et al.'s version \cite{urbanNominalRec,DBLP:journals/jar/Urban08} implemented in Isabelle/Nominal is an improvement of Pitts's in that it %(tacitly) 
removes the Finite Support requirement from the models---which is practically %quite 
significant because it enables non-finitely supported target domains for recursion. 
%(Pitts may have been neglecting it due to the desire to stay within a universe of nominal sets.)
%
Norrish's result \cite{primrecFOAS-Norrish04} is explicitly inspired by nominal logic, 
%for its underlying intuitions,  
but renounces the definability of the free-variable operator 
from swapping---with the price of taking both swapping and free-variables as primitives. 
My previous work with Gunter and Gheri takes as primitives either term-for-variable substitution and freshness \cite{DBLP:conf/icfp/PopescuG11} or swapping and freshness
\cite{DBLP:journals/jar/GheriP20}, and requires properties expressed by different Horn clauses (and does not explore a Barendregt dimension, like Pitts, Urban et al.\ and Norrish do).   
%
%In particular, 
My previous focus on term-for-variable substitution \cite{DBLP:conf/icfp/PopescuG11} (as opposed to renaming, i.e., variable-for-variable substitution) impairs expressiveness---for example, the depth %(length) 
of a term is not definable using a recursor based on term-for-variable substitution because we cannot say how term-for-variable substitution affects the depth of a term based on its 
depth and that of the substitutee alone. 
My current result based on rensets keeps freshness out of the  primitive operators base (like nominal logic does), and provides an unconditionally equational characterization using only  constructors and renaming. The key to achieving this  
%level of purging 
minimality is the simple expression of freshness from renaming in my axiomatization of rensets. 
In future work, I plan %to pursue 
a systematic formal comparison of the relative expressiveness 
of all these nameful recursors. 
\looseness=-1

%\subsection
\vspace*{1ex}
\noindent
\textbf{Recursors in other paradigms.} 
%
%As expected, 
Fig.~\ref{fig-relatedWork} focuses on nameful recursors, while only the Fiore et al.\ / Hofmann  recursor 
for the sake of a rough comparison with the nameless approach. 
I should stress that such a comparison is necessarily rough, since the nameless recursors do not give the same ``payload'' as the nameful ones. 
This is because of the handling of bound variables. In the nameless paradigm, the $\lambda$-constructor does not explicitly take a variable as an input, as in $\Lm\;x\;t$, i.e., does not have type $\Var \ra \Trm \ra \Trm$.
Instead, the bindings are indicated
 through nameless pointers to %binding 
 positions in a term. 
So the nameless $\lambda$-constructor, let's call it $\DBLm$, takes only a term, as in $\DBLm\;t$, i.e., has type 
$\Trm \ra \Trm$ or a scope-safe (polymorphic or dependently-typed) variation of 
this, e.g., $\prod_{n \in \mathbb{F}} \Trm_n \ra \Trm_{n+1}$ %as in 
\cite{fio-abs,DBLP:conf/lics/Hofmann99} 
or 
$\prod_{\alpha \in \Type} \Trm_\alpha \ra \Trm_{\alpha + \unit}$ %as in 
 \cite{BirdP-lambda,DBLP:conf/csl/AltenkirchR99}.  
The $\lambda$-constructor is of course matched by operators in the considered models, which appears in the clauses of the functions $f$ defined recursively on terms: Instead of a clause of the form 
$f\;(\Lm\;x\;t) \,=\, \< \mbox{expression depending on $x$ and $f\,t$}\>
$
from the nameful paradigm, 
in the nameless paradigm one gets a clause of the form
$
f\;(\DBLm\;t) \,=\, \< \mbox{expression depending on $f\,t$}\>
$. 
A nameless recursor is usually easier to prove correct and easier to apply because the nameless constructor $\DBLm$ is free---whereas % By contrast, %as already discussed, 
a nameful recursor must wrestle with the non-freeness of $\Lm$, 
%the issue of the definition being independent  from the choice of the bound  variable hence overall compatible with alpha-renaming, 
%which is 
handled by verifying certain properties of the target models.  However, once the definition is done, having nameful clauses pays off by allowing ``textbook-style'' proofs that stay close to the informal presentation of a calculus or 
logic, whereas with the nameless definition some additional index shifting bureaucracy is necessary.  (See \cite{DBLP:journals/entcs/BerghoferU07} for a  detailed discussion, and \cite{locallyNameless} for a hybrid solution.)

A comparison of nameful recursion with HOAS recursion %\cite{phe-primRecHOAS,DBLP:conf/esop/0001P17,beluga} is even more
is also generally difficult, since major HOAS frameworks such as 
Abella \cite{abellaJournalPaper}, Beluga \cite{beluga} or Twelf \cite{DBLP:conf/cade/PfenningS99}
are developed within non-standard logical foundations, allowing a $\lambda$-constructor of type $(\Trm \ra \Trm) \ra \Trm$, which is not amenable to typical well-foundedness based recursion but requires some %more exotic 
custom solutions (e.g., \cite{DBLP:journals/tcs/SchurmannDP01,DBLP:conf/esop/0001P17}).  
However, the \emph{weak HOAS} variant 
\cite{weakHOAS,gun-proper} employs a  
constructor of the form $\WHLm : (\Var \ra \Trm) \ra \Trm$ which \emph{is} recursable, and in fact yields a free datatype, let us call it $\WHTrm$---one generated by $\WHVr : \Var \ra \WHTrm$, $\WHAp : \WHTrm \ra \WHTrm \ra \WHTrm$ and $\WHLm$. 
%, the last being essentially an infinitary constructor. 
$\WHTrm$ contains (natural encodings of) all terms but also additional entities referred to as ``exotic terms''.  Partly because of the exotic terms, this free datatype by itself is not 
very %particularly 
helpful for recursively defining useful functions on terms. But 
the situation is dramatically improved if one employs a variant of weak HOAS called \emph{parametric HOAS (PHOAS)} \cite{chlipalaParamHOAS2008}, i.e., takes $\Var$ not as a fixed type but as a type parameter (type variable)  and works with $\prod_{\Var \in \Type} \Trm_\Var$; this enables many useful definitions by choosing a suitable type $\Var$ (usually large enough to make the necessary distinctions) and then performing standard recursion. The functions definable in the style of PHOAS seem to be exactly those definable via the semantic domain interpretation pattern %at which my renaming-based recursor excels 
(Section~\ref{subsec-semInt}): Choosing the instantiation of $\Var$ to a type $T$ corresponds to employing environments in $\Var \ra T$. (I illustrate this at the end of Appendix~\ref{app-moreRecExa} 
by showing the semantic-domain version of a PHOAS example.)
%$\eta$-reducibility testing example in the PHOAS paper \cite[Fig.3]{chlipalaParamHOAS2008}.  
%

As a hybrid nameful/HOAS approach we can count Gordon and Melham's characterization of the datatype of terms \cite{DBLP:conf/tphol/GordonM96}, which employs the nameful constructors but formulates recursion treating $\Lm$ as if recursing in the 
weak-HOAS datatype $\WHTrm$. Norrish's %already discussed 
recursor \cite{primrecFOAS-Norrish04} (%shown 
a participant in Fig.~\ref{fig-relatedWork}) has been inferred from Gordon and Melham's one. 
Weak-HOAS recursion also has interesting connections with nameless recursion: In presheaf toposes such as those employed by Fiore et al.~\cite{fio-abs}, Hofmann \cite{DBLP:conf/lics/Hofmann99} and Ambler et al.~\cite{DBLP:conf/icfp/AmblerCM03}, 
for any object $T$ the function space $\Var \Ra T$ is isomorphic 
to the De Bruijn level shifting transformation applied to $T$; this effectively equates 
%,
%this identifies, %in that setting, % the weak-HOAS and nameless $\lambda$-constructors, hence 
the weak-HOAS and nameless recursors.  
A final cross-paradigm note: %While especially successful in the nameful paradigm, 
In themselves, nominal sets are not confined to the nameful paradigm; 
their category is equivalent \cite{DBLP:conf/lics/GabbayP99} to the Schanuel topos \cite{johnstone-shanuelTopos}, %an attractive category 
which is attractive for pursuing the nameless approach. 

\vspace*{1ex}
\noindent 
\textbf{%Algebraic axiomatizations and model theory for syntax with bindings.}
Axiomatizations of renaming.} 
In his study of name-passing process calculi, Staton \cite{statonThesis} considers an enrichment of nominal sets with renaming (in addition to swapping) 
and axiomatizes renaming with the help of the nominal (swapping-defined) freshness predicate. He shows that the resulted category is equivalent to the non-injective renaming counterpart of the Schanuel topos (i.e., the subcategory of $\Set^{\mathbb{F}}$ consisting of functors that preserve pullbacks of monos).  
Gabbay and Hofmann 
\cite{gabbayHofmann-nominalRenamingSets} provide an elementary characterization of the above category, in terms of \emph{nominal renaming sets}, 
%\footnote{I am indebted to one of the anonymous reviewers for pointing me to this work.} 
which are sets equipped with a multiple-variable-renaming action satisfying identity and composition laws, and a form of Finite Support (FS). Nominal renaming sets seem very related to rensets satisfying FS. Indeed, any nominal renaming set forms a FS-satisfying renset when restricted to single-variable renaming. Conversely, I conjecture that any FS-satisfying renset gives rise to a nominal renaming set. %, but I was not able to prove this yet. 
This correspondence seems similar to the one between the permutation-based and swapping-based  alternative axiomatizations of nominal sets---in that the two express the same concept up to an isomorphism of categories. In their paper, Gabbay and Hofmann  do not study renaming-based recursion, beyond noting the availability of a recursor stemming from the functor-category view (which, as I discussed above, enables nameless recursion with a weak-HOAS flavor).   Pitts \cite{pittsNominalVsCubical} introduces \emph{nominal sets with $01$-substitution structure}, which axiomatize substitution of one of two possible constants for variables on top of the nominal axiomatization, and proves that they form a category that is equivalent with that of cubical sets \cite{DBLP:conf/types/BezemCH13}, hence relevant for the univalent foundations 
%homotopy type theory 
\cite{hottbook}.  

%\subsection
\vspace*{1ex}
\noindent
\textbf{%Algebraic axiomatizations and model theory for syntax with bindings.}
Other work.} 
Sun \cite{sun-alg} develops universal algebra for first-order languages with bindings (generalizing %previous 
work by Aczel \cite{acz-fre}) and proves a completeness theorem. In joint  work with Ro\c{s}u \cite{pop-TGL}, I develop first-order logic and prove completeness on top of a generic syntax with axiomatized free-variables and substitution. 
%

%\subsection
\vspace*{1ex}
\noindent
\textbf{Renaming versus swapping and nominal logic, final round.} 
%
%
%I have made comparisons with nominal sets all throughout the paper, highlighting the benefits of my renaming-based alternative.  
%However, with regard to formal technology for syntax with bindings, 
I believe that my work complements rather than competes with nominal logic. 
%
%todo: move to related work 
%Thus, 
My results do not challenge the swapping-based approach to defining syntax (defining the alpha-equivalence on pre-terms and quotienting to obtain terms)  
recommended by nominal logic, which is more elegant than a renaming-based alternative; 
%still seems like the most elegant approach. On the other hand, my easier-to-apply recursor could be a useful addition on top of the nominal substratum. 
%syntactic renaming is more difficult to define than syntactic swapping---which means that the swapping-based approach to defining syntax
%bootstrapping the syntax (defining the alpha-equivalence on pre-terms and quotienting to obtain terms etc.) 
%recommended by nominal logic still seems like the most elegant approach. 
but my easier-to-apply recursor can be a useful addition even on top of the nominal substratum. 
%
%On the other hand, as I have argued in the paper, rensets seem to provide improvements on the recursion front---which suggest that renaming-based recursion might be a useful principle to provide via definitional packages such as Isabelle/Nominal or my bindings-as-functors package (under construcion). 
Moreover, some of my constructions are explicitly inspired by the nominal ones. For example, I started by adapting the nominal idea of 
defining freshness from swapping before noticing that renaming enables %instead 
%an even 
a simpler %(equational) 
formulation. % (Prop.~\ref{prop-freshIff}). 
My formal treatment of Barendregt's variable convention also originates from nominal logic---as it turns out, %the Barendregt parameter set can be ``plugged in'' 
this idea works equally well in my setting.
In fact, %suspect 
%conjecture  
I came to believe that the 
%it seems that the 
possibility of a Barendregt enhancement %of a principle 
is largely orthogonal to the particularities of a binding-aware recursor. 
 In future work, I plan to investigate this, i.e., seek general conditions under which an initiality principle (such as Theorems~\ref{thm-nominalnit} and \ref{thm-termInit}) is amenable to a Barendregt enhancement (such as Theorems~\ref{thm-pittsRec} and \ref{thm-substRec}, respectively).

%\subsection{Other work}

%Staton \cite[Chapter 7]{statonThesis}, in the context of analysing name-passing calculi, introduces nominal rensets---but these are defined as a theory in nominal logic; so permutatiion and freshness are pre-requisits. 

\ \\
{\bf Acknowledgments. }
%\small
%
I am grateful to the IJCAR reviewers for their insightful comments and suggestions, and for pointing out related work. 
%

%%%%%%%%%%%%%%%%
\bibliographystyle{splncs03}
\bibliography{bib}

\include{SSappendix}

\end{document}

%% file: myCommandsSubstSets.tex
\usepackage{amssymb}

\newenvironment{myitem}[1][]
{\itemize[leftmargin=4.5ex,topsep=1ex,itemsep=1pt, #1]}
{\enditemize}
\newenvironment{myitemm}[1][]
{\itemize[leftmargin=6ex,topsep=1ex,itemsep=1pt, #1]}
{\enditemize}
\newenvironment{myitemmm}[1][]
{\itemize[leftmargin=22ex,topsep=1ex,itemsep=1pt, #1]}
{\enditemize}
\newenvironment{myitemmmm}[1][]
{\itemize[leftmargin=20ex,topsep=1ex,itemsep=1pt, #1]}
{\enditemize}

\newtheorem{mylemma}{Lemma}
 
\newtheorem{prop}[mylemma]{Prop}
\newtheorem{thm}[mylemma]{Theorem}

\newcommand{\leftOut}[1]{}

\newbox\boxA

\newcommand{\oexp}{\mbox{\hphantom{$+_{\mathsf{o}}$}}\llap{$\text{\textasciicircum}_{\mathsf{o}}\kern.1em$}}
\newcommand\hlt[1]{\mbox{\colorbox{light-gray}{$#1$}}}

\newcommand{\cexp}{\mbox{\hphantom{$+_{\mathsf{o}}$}}\llap{$\text{\textasciicircum}_{\mathsf{c}}\kern.1em$}}
\newcommand{\imageOp}{\raise.2ex\hbox{\mbox{$\scriptscriptstyle\bullet$}}}
\newcommand{\vimageOp}{\raise.2ex\hbox{\mbox{$\scriptscriptstyle-\hspace*{-0.7ex}-\kern-.2em\bullet$}}}

\newcommand\TC{\sf}

\newcommand\CHOPFROMUN{.25}
\newcommand\UN{{\setbox\boxA=\hbox{\_}\usebox\boxA\kern-\CHOPFROMUN\wd\boxA{\color{white}\vrule height 0ex depth .444ex width \CHOPFROMUN\wd\boxA}\kern-\CHOPFROMUN\wd\boxA}}
\newcommand{\sm}{\smallsetminus}

\renewcommand{\phi}{\varphi}

\newcommand\wgeq\gtrsim

\newcommand{\su}{\subseteq}
\renewcommand{\iff}{\allowbreak\mathrel{\;\Leftarrow\nobreak\kern-1.6ex\Rightarrow\;}\allowbreak} %% TYPESETTING

\newcommand{\ra}{\rightarrow}
\newcommand{\Ra}{\Rightarrow}
\newcommand{\lra}{\rightarrow}
\newcommand{\llra}{\leftrightarrow}

\newcommand{\llam}{{\llam}}

\renewcommand\[{[\hspace*{-0.35ex}[}  
\renewcommand\]{]\hspace*{-0.35ex}]}  
\newcommand\alb\allowbreak
\newcommand{\<}{\langle}
\renewcommand{\>}{\rangle}

\newcommand{\Nom}{\underline{\mathit{Nom}}}

\newcommand{\Sbs}{\underline{\mathit{Sbs}}}
\newcommand{\FSbs}{\underline{\mathit{FSbs}}}
\newcommand{\NomCE}{\underline{\mathit{Nom}}_{\mathsf{CE}}}
\newcommand{\SbsCE}{\underline{\mathit{Sbs}}_{\mathsf{CE}}}
\newcommand{\SSbsCE}{\underline{\mathit{SSbs}}_{\mathsf{CE}}}

\renewcommand{\SS}{\mathcal{I}}
\renewcommand{\AA}{\mathcal{A}}

\newcommand{\TTrm}{\mathcal{T}\hspace*{-0.3ex}rm}

\newcommand{\BB}{\mathcal{B}}

\newcommand{\FF}{\mathcal{F}}

\newcommand\SConnect{\mathsf{Con}_{\tiny \_[\_/\!\_]}}

\newcommand\ap{{\mathsf{ap}}}
\newcommand\lm{{\mathsf{lm}}}

\newcommand\clam{{\mathsf{clam}}}
\newcommand\cfv{{\mathsf{cfv}}}
\newcommand\cbv{{\mathsf{cbv}}}
\newcommand\cbvs{{\mathsf{cbvs}}}
\newcommand\sem{{\mathsf{sem}}}
\newcommand\canEta{{\mathsf{canEta}}}

\newcommand{\Vr}{\mathsf{{Vr}}}
\newcommand{\WHVr}{\mathsf{{WHVr}}}
\newcommand{\Lm}{\mathsf{{Lm}}}
\newcommand{\DBLm}{\mathsf{{NLm}}}

\newcommand{\WHLm}{\mathsf{{WHLm}}}
\newcommand{\Ap}{\mathsf{{Ap}}}
\newcommand{\WHAp}{\mathsf{{WHAp}}}

\newcommand{\PVr}{\mathsf{{PVr}}}
\newcommand{\PLm}{\mathsf{{PLm}}}
\newcommand{\PAp}{\mathsf{{PAp}}}

\newcommand{\fresh}{\#}

\newcommand{\Term}{{{\mathsf{Term}}}}

\newcommand{\llength}{{{\mathsf{length}}}}

\newcommand{\FV}{{{\mathsf{FV}}}}

\newcommand{\Pow}{{{\mathsf{Pow}}}}

\newcommand{\supp}{{{\mathsf{supp}}}}

\renewcommand{\max}{{{\mathsf{max}}}}

\newcommand{\ttrue}{\mathsf{{true}}} 

\newcommand{\ffalse}{\mathsf{{false}}}  
\newcommand{\Var}{\mathsf{{Var}}} 
 
\newcommand{\Type}{\mathsf{{Type}}}

\newcommand{\Trm}{\mathsf{Trm}}
\newcommand{\WHTrm}{\mathsf{WHTrm}}
\newcommand{\PTrm}{\mathsf{PTrm}}

\newcommand{\Perm}{\mathsf{{Perm}}}

\newcommand{\id}{\mathsf{{id}}}

\newcommand\sww{\_[\_\!\hspace*{0.2ex}\sw\hspace*{0.2ex}\!\_]}
\newcommand{\sw}{\hspace*{-0.25ex}\wedge\hspace*{-0.20ex}}

\newcommand{\Bool}{{\TC Bool}}

\newcommand{\Nat}{\mathbb{N}}

\newcommand{\Set}{{\TC \mathit{Set}}}

\newcommand{\unit}{{\TC unit}}

\newmuskip\originalthinmuskip
\originalthinmuskip=\thinmuskip
\newmuskip\tinyGapMu
\renewcommand\ldots{\mathinner{.\mskip\originalthinmuskip .\mskip\originalthinmuskip .}}

\newcommand\Smash[1]{\kern-200mm\smash{#1}\kern-200mm}
\newcommand\SubItem[2]{\indent\hbox to \leftmargini{\hfill#1\enskip}#2}

\newcommand\XDot{\raise1ex\hbox{\Large.\kern.1em}}

\def\figurename{Fig\hbox{.}}

%% file: SSappendix.tex
\appendix

\begin{center}
	{\Large APPENDIX}
\end{center}

\section{More Example Functions Definable with the Renaming-Based Recursor}
\label{app-moreRecExa} 

For each of the following examples, I will not indicate the required CE renset or BCE renset. Rather, we show the clauses describing the behavior of the defined function with respect to the constructors and (variable-for-variable) substitution, from which the corresponding structure on the target domain can be easily inferred (like we did in Section~\ref{subsec-semInt}). In each case, the verification of the necessary properties to deploy our Theorem~\ref{thm-substRec} is trivial. 

\ \\
%\subsection
\noindent
{1. The length of a term \cite[Example 4.2]{pitts-AlphaStructural}, } 
$\llength : \Term \ra \Nat$. 

\begin{myitem}
	\item $\llength\;(\Vr\;x) = 1$
	\item $\llength\;(\Ap\;t_1\;t_2) = 
	\max\,(\llength\;t_1,\llength\;t_2) + 1$
	\item $\llength\;(\Lm\;x\;t) = \llength\;t + 1$
	\item $\llength\;(t\,[x\sw y]) = \llength\;t$ 
\end{myitem}

%\ \\
%\subsection
\noindent
{2. Counting $\lambda$-abstractions \cite[Example 8.18]{pitts_2013}, } 
$\clam : \Term \ra \Nat$. 
%(very smilar to the previous one). 

\begin{myitem}
	\item $\clam\;(\Vr\;x) = 0$
	\item $\clam\;(\Ap\;t_1\;t_2) = 
	\clam\;t_1 + \clam\;t_2$
	\item $\clam\;(\Lm\;x\;t) = \clam\;t + 1$
	\item $\clam\;(t\,[x\sw y]) = \clam\;t$ 
\end{myitem}

%\ \\
%\subsection
\noindent
{3. Counting the number of free occurrences of a variable, } 
$\cfv : \Term \ra \Var \ra \Nat$. 

\begin{myitem}
	\item 
$\cfv\ (\Vr\ y)\ x= \mbox{(if $x = y$ then $1$ else 0)}$
\item 
$\cfv\ (\Ap\ t_1\ t_2)\ x= \cfv\ t_1\ x + \cfv\ t_2\ x$ 
\item 
$	\cfv\ (\Lm\ y\ t)\ x = 
\mbox{(if $x = y$ then $0$ else $\cfv\ t\ x$)}$
\item $\cfv\ (t[z / y])\ x =  
\left\{\begin{array}{l}
  \cfv\;t\;x \mbox{ ,  if $x \notin\{y,z\}$}
   \\
  \cfv\;t\;x + \cfv\;t\;y  \mbox{ ,  if $x = z \not= y$}
  \\
  0 \mbox{ ,  if $x = y \not= z$}
  \\
  \cfv\;t\;y \mbox{ ,  if $x = y = z$}
\end{array}\right.$
\end{myitem} 
%

%\ \\
%\subsection
\noindent
{4. Term-for-variable substitution \cite[Example 8.18]{pitts_2013}, } 
$\_\[\_\,/\_\] : \Trm \ra \Trm \ra \Var \ra \Trm$. 

\begin{myitem}
	\item $(\Vr\;y)\,\[s/x\] = 
	\left\{\begin{array}{l}
	s\mbox{ ,  if $x=y$}
	\\
	\Vr\;y \mbox{ ,  otherwise}
	\end{array}\right.$
\item  $(\Ap\;t_1\;t_2)\,\[s/x\] = \Ap\,(t_1\,\[s/x\])\,(t_2\,\[s/x\])$
\item $(\Lm\;y\;t)\,\[s/x\] = \Lm\;y\,(t\,\[s/x\])$ if $y \notin \{x\} \cup \FV\,s$ 
\item $t\,[y/z]\,\[s/x\] = t\,\[s/x\]\,[y/z]$ if $y,z \notin \{x\} \cup \FV\,s$ 
\end{myitem} 
(So here one applies the recursion principle %(Theorem~\ref{thm-substRec})
 with the Barendregt parameter $X$ taken to be $\{x\} \cup \FV\,s$.) 

\ \\
%\subsection
\noindent
{5. Defining (capture-avoiding) parallel term-form-variable substitution, } 
$\_\[\_\] : \Trm \ra (\Var \ra_{\textsf{\small fin}} \Trm) \ra \Var \ra \Trm$, 
where $\Var \ra_{\textsf{\small fin}} \Trm$ is the set of functions $\rho :\Var\ra\Trm$ having $\supp\;\rho$ finite (where $\supp\,\rho$ consists of all variables $x$ such that $\rho\;x \not=\Vr\;s$), is similar: 
\begin{myitem}
	\item $(\Vr\;y)\,\[\rho\] = \rho\,y$
	\item  $(\Ap\;t_1\;t_2)\,\[\rho\] = \Ap\,(t_1\,\[\rho\])\,(t_2\,\[\rho\])$
	\item $(\Lm\;y\;t)\,\[\rho\] = \Lm\;y\,(t\,\[\rho\])$ if $y \notin \supp\;\rho$ 
	\item $t\,[y/z]\,\[\rho\] = t\,\[\rho\]\,[y/z]$ if 
	$y,z \notin \supp\;\rho$ and $y,z \notin \FV\,(\rho\;x)$ 
	for all $x\in\Var$.   
\end{myitem} 
(So here one applies the recursion principle %(Theorem~\ref{thm-substRec})
with the Barendregt parameter $X$ taken to be $\supp\;\rho \cup \{\FV\,(\rho\;x) \mid x \in \supp\;\rho\}$.)

\ \\
%\subsection
\noindent
{6. Counting the number of bound variables of a term \cite{DBLP:journals/tcs/SchurmannDP01,pitts_2013}},  $\cbv : \Trm \ra \Nat$. 
Note that the notion of semantic domain can be chosen flexibly, to also cover certain purely syntactic operators as well.  A particular case is the function $\cbv$ that counts the number of bound variables of term. In his book \cite{pitts_2013}, Pitts defines it following the approach of Sch\"{u}rmann et al.\ \cite{DBLP:journals/tcs/SchurmannDP01} as $\cbv \;t = \cbvs\;t\;(x \mapsto 0)$, where the auxiliary function $\cbvs : \Trm \ra (\Var \ra \Nat) \ra \Nat$ operates according the following recursive clauses:
\begin{myitem}
	\item[(1)] $\cbvs\,(\Vr\;x)\,\xi = \xi\;x$
	\item[(2)]  $\cbvs\,(\Ap\;t_1\,t_2)\,\xi = (\cbvs\;t_1\,\xi) + (\cbvs\;t_2\,\xi)$
	\item[(3)]  $\cbvs\,(\Lm\;x\;t)\,\xi = \cbvs\;t\,(\xi\<x:= 1\>)$
\end{myitem}
(The same complications with deploying the nominal recursor arise, and Pitts deploys a similar workaround.)

\ \\
%\subsection
\noindent
{6. The $\eta$-reducibility testing example from the PHOAS paper \cite[Fig.3]{chlipalaParamHOAS2008}, } 
$\canEta : \Term \ra \Bool$ is defined by 

$$\canEta\;t = 
\left\{\begin{array}{l}
	\ttrue\mbox{ ,  if $t$ has the form $\Lm\;x\;(\Ap\;s\;x)$}
	\\
	\mbox{\phantom{$\ttrue$ ,}
		and $\canEta'\;s\;(\top\<x:= \ffalse\>) = \ttrue$}
	\\
	\ffalse \mbox{ ,  otherwise}
\end{array}\right.$$
\noindent 	
where $\top$ is the environment sending all variables to $\ttrue$ and 
$\canEta' : \Term \ra (\Var \ra \Bool) \ra \Bool$ is defined as an instance of the semantic domain interpretation pattern described in Section~\ref{subsec-semInt}: 
\begin{myitem}
	\item[(1)] $\canEta'\,(\Vr\;x)\,\xi = \xi\;x$
	\item[(2)]  $\canEta'\,(\Ap\;t_1\,t_2)\,\xi = (\canEta'\;t_1\,\xi) \,\&\, (\canEta'\;t_2\,\xi)$
	\item[(3)]  $\canEta'\,(\Lm\;x\;t)\,\xi = \canEta'\;t\,(\xi\<x:= \ttrue\>)$
\end{myitem}
\noindent 
($\canEta$ can of course be alternatively defined by other means, e.g., using the free-variable operator.)

\section{Full-Fledged Primitive Recursion}
\label{app-fullFledgedRec}

Theorem~\ref{thm-substRec} restricts the recursive behavior 
to iteration, which allows the value of the defined function $f$ on a term $t$ to depend on the value of $f$ on the components of $t$. For example, if $t$ has the form $\Ap\,t_1\,t_2$, then $f\,t$ can depend on $f\;t_1$ and $f\;t_2$. This can routinely be extended to full primitive recursion, which additionally allows the dependence on the components themselves (not necessarily through $f$),  e.g., on $t_1$ and $t_2$. All the recursors recalled in Fig.~\ref{fig-relatedWork} of the main paper admit full primitive recursion enhancements. My Theorem~\ref{thm-substRec} is no exception---here is its enhancement, where I highlighted the additions:

A \emph{full-recursion constructor-enriched substitutive set} (\emph{FRCE renset} for short) is a tuple $\AA = (A,\_[\_/\!\_],\Vr^\AA,\Ap^\AA,\Lm^\AA)$ where:
\begin{myitem}
	\item  $(A,\_[\_/\!\_])$ is a substitutive set
	\item  $\Vr^\AA : \Var \ra A$, 
	$\Ap^\AA : \hlt{\Trm \ra } A \ra \hlt{\Trm \ra }  A \ra A$, 
	$\Lm^\AA : \Var \ra \hlt{\Trm \ra }  A \ra A$ are operators on $A$ with arities matching those of the term constructors
\end{myitem}
such that the following properties hold for all $x,y,z\in \Var \sm X$, $\hlt{t,t_1,t_2\in \Trm}$ and $a,a_1,a_2\in \Trm$: 
\begin{description}
	\item [\mbox{\normalfont{(RS1)}}] $(\Vr^\AA\;x)[y/z] = \Vr^\AA (x[y/z])$
	\item [\mbox{\normalfont{(RS2)}}] $(\Ap^\AA\hlt{t_1} a_1 \hlt{t_1} a_2)[y/z] = \Ap^\AA\hlt{(t_1[y/z])} (a_1[y/z]) \hlt{(t_2[y/z])} (a_2[y/z])$
	\item [\mbox{\normalfont{(RS3)}}]  if $x\notin \{y,z\}$ then 
	$(\Lm^\AA x \hlt{t} a)[y/z] = \Lm^\AA x \hlt{(t[y/z])} (a[y/z])$
	
	\item [\mbox{\normalfont{(RS4)}}]  $(\Lm^\AA x\hlt{t} a)[y/x] = \Lm^\AA\,x \hlt{t}  a$
	
	\item [\mbox{\normalfont{(RS5)}}]   if $z\not=y$ then $\Lm^\AA x\hlt{(t[z/y])}(a[z/y]) = \Lm^\AA y \hlt{(t[z/y][y/x])} (a[z/y][y/x])$

\end{description}
%
%We will again call $\Vr^\AA,\Ap^\AA,\Lm^\AA$ the \emph{constructors} of $\AA$.

\begin{thm} \rm  
	\label{thm-substRecFullPrimrec}
	Let $X$ be a finite set 
	and $\AA = (A,\_[\_/\!\_],\Vr^\AA,\Ap^\AA,\Lm^\AA)$ be a FRCE renset.  %	
	Then there exists a unique function $f: \Trm \ra A$ such that the following hold: 
	\begin{myitem}
		\item[(i)]  $f\,(\Vr\;x) = \Vr^\AA\;x$
		\item[(ii)]  $f\,(\Ap\;t_1\;t_2) = \Ap^\AA\hlt{t_1}(f\;t_1)\hlt{t_2}(f\;t_2)$
		\item[(iii)] $f\,(\Lm\;x\;t) = \Lm^\AA\;x\hlt{t}(f\;t)$ if $x\notin X$
		\item[(iv)] $f\,(t[y/z]) = (f\;t)[y/z]$ if $y,z\notin X$
	\end{myitem}
\end{thm}

%%%
\section{Alternative Definition of Nominal Sets}
\label{app-swapVsPermNominal}

In the main paper, I focused on the swapping-based presentation of nominal sets. Here I recall an equivalent formulation based on permutations. % and explain why, for all intents and purposes,  
The equivalence between the two formulations is described in detail in Pitts's monograph \cite[Section 6.1]{pitts_2013}. 

Let $\Perm$ denote the set of finite permutations (i.e., bijections of finite support) on the set of variables, 
namely 
$\{\sigma : \Var \ra \Var \mid 
\{x \mid \sigma\;x \not=x\}$ finite $\}$. Let $\id$ denote the identity permutation. 
Let $x \llra y$ denote the $(x,y)$-transposition, i.e., the 
permutation that takes $x$ to $y$, $y$ to $x$ and every other variable to itself. 

Let us call \emph{$\Perm$-nominal set} any pair $(A,\_[\_])$ where 
$A$ is a set and  $\_[\_] : A \ra \Perm \ra A$
is an action on $A$ of the permutation group $\Perm$, i.e.,  
it satisfies the following properties for all $a\in A$ and $\sigma,\sigma'\in\Perm$: 
\begin{myitem} 
\item{Identity: \ } $a[\id] = a$
\item{Compositionality: \ } 
$a[\sigma] [\sigma']= a[\sigma \circ \sigma']$ 
\end{myitem} 

Given two $\Perm$-nominal sets $\AA = (A,\_[\_])$ and $\BB = (B,\_[\_])$, a \emph{$\Perm$-nominal morphism} $f:\AA \ra \BB$ is a function $f:A\ra B$ that commutes with the permutation operation, in that $(f\;a)[\sigma] = f(a[\sigma])$ for all $a\in A$ and $\sigma\in\Perm$. $\Perm$-nominal sets and $\Perm$-nominal morphisms form a category denoted by $\Nom_\Perm$.  

Let $\AA = (A,\_[\_])$ be a $\Perm$-nominal set. 
The notion of finite support of an element $a\in A$ is defined similarly to that from nominal sets (described in Section~\ref{subsec-nominal}), but using $a[x\llra b]$ rather than $a[x \sw y]$.  In fact, it is easy to see that $A$ together with the restriction of $\_[\_]$ to transpositions forms a nominal set. Let $G\,\AA$ denote this nominal set. The notions of support and freshness in $\AA$ coincide with those in $\AA$; and any function $f: A \ra B$ is a $\Perm$-nominal morphism between $\AA$ and $\BB$ iff it is a nominal morphism between $G\,\AA$ and $G\,\BB$. 

Conversely, let $\AA = (A,\_[\_\!\sw\!\_])$ be a nominal set. One can extend the swapping operator $\_[\_\!\sw\!\_] : A \ra \Var \ra \Var  \ra A$ to a permutation operator 
$\_[\_] : A \ra \Perm \ra A$ by taking advantage of the fact that any permutation $\sigma$ is decomposable into a sequence of transpositions, 
$\sigma = (x_1\llra y_1) \cdot \ldots \cdot (x_n \llra y_n)$. 
Namely, one defines $a[\sigma] = a[x_1\sw y_1]\ldots[x_n\sw y_n]$. 
%Since the decompostion is not unique, we need to prove that this is a correct definition, in that the lefthand side of the equality does not depend on the particular decomposition. And indeed, it turns out that the swappable-set axioms are able to ensure this property: 
Thanks to the nominal set properties, this definition is correct, in that it does not depend on representatives. Moreover, $A$ together with the just defined $\_[\_]$ forms a $\Perm$-nominal set. Let $H\,\AA$ denote this $\Perm$-nominal set. Again, the notions of support and freshness in $\AA$ coincide with those in $\AA$; and any function $f: A \ra B$ is a nominal morphism between $\AA$ and $\BB$ iff it is a $\Perm$-nominal morphism between $H\,\AA$ and $H\,\BB$. 

Finally, the operators $G$ and $H$ are mutually inverse, and become mutually inverse functors when extended to be the identity on morphisms. 
The situation is summarized by the following theorem:

\begin{thm} \rm \label{thm-connections}  \cite[Theorem~6.1, Corollary 6.2]{pitts_2013}
	\\(1) The categories $\Nom$ and $\Nom_\Perm$ are isomorphic, via the functors $G: \Nom \ra \Nom_\Perm$ and $H: \Nom_\Perm \ra \Nom$ that are inverse to each other. 
	\\(2) The functors $G$ and $H$ preserve the carrier set and the notions of support and freshness.  
\end{thm}

The above means that, for all intents and purposes (at least as far as my results in this paper are concerned), the two definitions of nominal sets (namely nominal sets and $\Perm$-nominal sets) are equivalent. 

%We define 
%the permutation-action $\_[\_] : \Trm \ra \Perm \ra \Trm$, 
%\\e.g., we have %\\
%$(\Lm\;x\;(\Ap\;z\;y))\,[x \mapsto y,y\mapsto z,z\mapsto x] = 
%\Lm\;y\;(\Ap\;x\;z)$
%

\section{Substitutive Sets}
\label{app-SSsets}

Recall that the substitution operator I consider in the main paper is renaming, i.e., variable-for-variable substitution, 
$\_[\_/\!\_] : \Term \ra \hlt{\Var} \ra \Var \ra \Trm$. I will continue to refer to this as \emph{renaming}, 
while calling the term-for-variable substitution \emph{substitution} and denoting it by 
$\_\[\_/\!\_\] : \Term \ra \hlt{\Trm} \ra \Var \ra \Trm$. 

I can use my characterization of the datatype of terms by means of renaming to obtain a characterization employing substitution. This does not seem to add practical value as a recursor, but may be of some theoretical interest as an equational characterization of terms by means of the constructors and substitution only. 

A \emph{substitutive set} is a triple $\AA = (A,\_\[\_/\!\_\],\Vr^\AA)$  where  
$\_\[\_/\!\_\] : A \ra A \ra \Var  \ra A$ 
and 
$\Vr^\AA : \Var \ra A$ are 
operators that satisfy the following properties, where I write $\_\[x/y\]$ instead of $\_\[(\Vr^\AA x)/y\]$
\begin{description}
	\item{Identity: \ } $a\[x/x\] = a$
	\item{Idempotence: \ } If $x\not=y \not= z$ then 
	$a\[(b\[z/y\])/y\]\[x/y\] = a\[(b\[z/y\])/y\]$
	\item{Chaining: \ } If $y\not=x_2$ then $a\[y/x_2\]\[x_2/x_1\] \[b/x_2\]= a\[y/x_2\]\[b/x_1\]$
	\item{Commutativity: \ } 
	If $y_2 \not= y_1 \not= x_1 \not= x_2$ then 
	\\
	$a\[(b\[y_2/y_1\])/x_1\] \[(c\[x_2/x_1\])/y_1\]= a\[(c\[x_2/x_1\])/y_1\]\[(b\[y_2/y_1\])/x_1\] $
\end{description}

Note the difference from the rensets discussed in the main paper: This time, since I axiomatize term-for-variable substitution, the second argument of the substitution operator is an element of the carrier set $A$ rather than a variable. However, there is also an ``embedding'' $\Vr^\AA$ of variables into $A$---this suggests a monadic structure on $A$, though the initial model characterization will not need such a requirement. 

Let $\AA = (A,\_\[\_/\!\_\],\Vr^\AA)$ be a substitutive set. 
If one define the freshness predicate similarly to how it was defined for substitutive sets (in Section~\ref{sec-substSet}), one obtains the following 
consequences of Idempotence, Chaining and Commutativity, which generalize natural properties connecting freshness with substitution for terms: 

\begin{prop} \rm
	The following hold in any substitutive set $\AA = (A,\_[\_],\Vr^\AA)$:
\begin{myitem} 
	\item %{Idempotence: \ }
	 If $x\not= y$ and $y\,\fresh\, b$ then 
$a\[b/y\]\[x/y\] = a\[b/y\]$
\item %{Chaining: \ } 
If $x_2 \,\fresh\,b$ then $a\[x_2/x_1\] \[b/x_2\]= a\[b/x_1\]$
\item %{Commutativity: \ } 
If $x \not= y$, $x \,\fresh\,a$ and $y\, \fresh\,b$ then 
$a\[b/x\] \[c/y\]= a\[c/y\]\[b/x\] $
\end{myitem} 
\end{prop}

A \emph{constructor-enriched substitutive set} (\emph{CE substitutive set} for short) is a tuple $\AA = (A,\_\[\_/\!\_\],\Vr^\AA,\Ap^\AA,\Lm^\AA)$ where:
\begin{myitem}
	\item  $(A,\_\[\_/\!\_\],\Vr^\AA)$ is a substitutive set
	\item  $\Ap^\AA : A \ra A \ra A$ and 
	$\Lm^\AA : \Var \ra A \ra A$ are operators on $A$ 
\end{myitem}
such that the following properties hold for all $a,a_1,a_2,b\in A$ and $x,y,z\in\Var$:
\begin{description}
	\item [\mbox{\normalfont{(S1)}}] $(\Vr^\AA x)\[b/y\] = $
	(if $x=y$ then $b$ else $\Vr^\AA x$)
	\item [\mbox{\normalfont{(S2)}}] $(\Ap^\AA a_1\;a_2)\[b/y\] = \Ap^\AA(a_1\[b/y\])\,(a_2\[b/y\])$
	\item [\mbox{\normalfont{(S3)}}]   if $x\not=z$ then 
	\\
	$(\Lm^\AA\;x\;a)\[(b\[z/x\])/y\] =$  (if $x=y$ then $\Lm^\AA\;x\;a$ else $\Lm^\AA x\,(a\[(b\[z/y\])/y\])$)
	\item [\mbox{\normalfont{(S4)}}]   if $z\not=y$ then $\Lm\;x\;(a\[z/y\]) = \Lm\;y\;(a\[z/y\]\[y/x\])$
\end{description}

The notion of \emph{CE substitutive morphism} is defined 
as the expected: 
as a function between two CE substitutive sets that commutes with both substitution and the constructors.  
Let $\SSbsCE$ be the category of CE substitutive sets and morphisms. 

\begin{thm} \rm \label{thm-strongInit}
	$(\Trm,\_\[\_/\!\_\],\Vr,\Ap,\Lm)$ 
	is the initial CE substitutive set, i.e., is the initial object in $\SSbsCE$. 
\end{thm}
\emph{Proof idea.} Let  $\AA = (A,\_\[\_/\!\_\],\Vr^\AA,\Ap^\AA,\Lm^\AA)$ be a substitutive set.  
Defining $\_[\_/\!\_] : A \ra \Var \ra \Var  \ra A$ by 
$a[y / z] = a [(\Vr^\AA y) / z]$ produces a substitutive set. Applying Theorem~\ref{thm-termInit}, 
one obtains a CE renset morphism 
$f:\Trm \ra A$. By fresh induction on terms and using the properties of 
substitutive sets, one then proves that $f$ commutes not only 
with renaming, but also with substitution---which makes it  
a CE substitutive morphism. Uniqueness holds immediately from 
the choice of $f$.  \qed